\documentclass[aps,letterpaper,twocolumn,prl,nofootinbib,floatfix]{revtex4}
\usepackage{amsmath}
\usepackage{amsfonts}
\usepackage{amssymb}
\usepackage[scanall]{psfrag}
\usepackage{psfrag}
\usepackage{graphicx}
\usepackage{color}

\begin{document}
\newcommand{\of}[1]{\left( #1 \right)}
\newcommand{\sqof}[1]{\left[ #1 \right]}
\newcommand{\abs}[1]{\left| #1 \right|}
\newcommand{\avg}[1]{\left< #1 \right>}
\newcommand{\cuof}[1]{\left \{ #1 \right \} }
\newcommand{\bra}[1]{\left < #1 \right | }
\newcommand{\ket}[1]{\left | #1 \right > }
\newcommand{\pil}{\frac{\pi}{L}}
\newcommand{\bx}{\mathbf{x}}
\newcommand{\by}{\mathbf{y}}
\newcommand{\bk}{\mathbf{k}}
\newcommand{\bp}{\mathbf{p}}
\newcommand{\bl}{\mathbf{l}}
\newcommand{\bq}{\mathbf{q}}
\newcommand{\bs}{\mathbf{s}}
\newcommand{\psibar}{\overline{\psi}}
\newcommand{\svec}{\overrightarrow{\sigma}}
\newcommand{\dvec}{\overrightarrow{\partial}}
\newcommand{\bA}{\mathbf{A}}
\newcommand{\bdelta}{\mathbf{\delta}}
\newcommand{\bK}{\mathbf{K}}
\newcommand{\bQ}{\mathbf{Q}}
\newcommand{\bG}{\mathbf{G}}
\newcommand{\bw}{\mathbf{w}}
\newcommand{\bL}{\mathbf{L}}
\newcommand{\ohat}{\widehat{O}}
\newcommand{\up}{\uparrow}
\newcommand{\down}{\downarrow}
\newcommand{\MM}{\mathcal{M}}
\newcommand{\tX}{\tilde{X}}
\newcommand{\tY}{\tilde{Y}}
\newcommand{\tZ}{\tilde{Z}}
\newcommand{\tOm}{\tilde{\Omega}}
\newcommand{\barA}{\bar{\alpha}}
\newcommand{\SN}{\mathcal{N}}

\author{E. Kapit$^{1,2}$}
\email{ekapit@mines.edu}

\author{P. Roushan$^{3}$}

\author{C. Neill$^{3}$}

\author{S. Boixo$^{4}$}

\author{V. Smelyanskiy$^{4}$}

\affiliation{$^{1}$Department of Physics, Colorado School of Mines, Golden, CO 80401, USA}
\affiliation{$^{2}$Department of Physics, Tulane University, New Orleans, LA 70118, USA}
\affiliation{$^{3}$Google, Inc, Santa Barbara, CA, USA}
\affiliation{$^{4}$Google, Inc, Venice, CA, USA}

\title{Entanglement and complexity of interacting qubits subject to asymmetric noise}

\begin{abstract}

The simulation complexity of predicting the time evolution of delocalized many-body quantum systems has attracted much recent interest, and simulations of such systems in real quantum hardware are promising routes to demonstrating a quantum advantage over classical machines. In these proposals, random noise is an obstacle that must be overcome for a faithful simulation, and a single error event can be enough to drive the system to a classically trivial state. We argue that this need not always be the case, and consider a modification to a leading quantum sampling problem-- time evolution in an interacting Bose-Hubbard chain of transmon qubits [Neill et al, Science 2018] -- where each site in the chain has a driven coupling to a lossy resonator and particle number is no longer conserved. The resulting quantum dynamics are complex and highly nontrivial. We argue that this problem is harder to simulate than the isolated chain, and that it can achieve volume-law entanglement even in the strong noise limit, likely persisting up to system sizes beyond the scope of classical simulation. Further, we show that the metrics which suggest classical intractability for the isolated chain point to similar conclusions in the noisy case. These results suggest that quantum sampling problems including nontrivial noise could be good candidates for demonstrating a quantum advantage in near-term hardware.

\end{abstract}

\maketitle

\section{Introduction}

Quantum sampling problems present the most promising near-term route to demonstrating ``quantum supremacy" \cite{preskill2011,harrowmontanaro2017}, where quantum hardware solves a problem that no classical supercomputer is capable of completing in a reasonable amount of time. Interest in these problems began with the boson sampling problem proposed by Aaronson and Arkhipov \cite{aaronsonarkhipov2013}, who argue that sampling the output distribution of groups of identical, noninteracting bosons propagating through a linear optical network is likely to be extremely difficult for classical machines. The years following that paper have seen a number of other candidate quantum systems put forward as challenging sampling problems \cite{bremner2011classical,boixoisakov2016,bremner2016average,miller2017quantum,bermejo2018architectures}, with perhaps the most attention focused on the random quantum circuit protocol \cite{boixoisakov2016}. This protocol is based on sampling the output of a random sequence of quantum gates acting on an initial product state, which is likely to be exponentially difficult for classical computers. Subsequent theoretical and experimental work \cite{neillroushan2017} extended this class of problems to include continuous time evolution (as opposed to a discrete collection of applied unitaries) in sampling the output of a time-evolving Bose-Hubbard chain, which like the other protocols is also very likely to be classically intractable once the system becomes sufficiently large. Since the threshold for superiority of quantum hardware depends on the state of the art in classical hardware and software, it naturally presents a moving target, and interest in quantum sampling problems has in turn prompted an explosion of progress in classical algorithms for simulating quantum circuits \cite{nevillesparrow2017,pednault2017breaking,boixo2017simulation,haner20170,chen201864,bouland2018quantum,chen2018classical,liwu2018,markov2018quantum,villalonga2018flexible,chen2019quantum,de2019massively}.

These sampling problems all involve simulating purely unitary quantum dynamics, and the introduction of local random noise into any of them reduces simulation fidelity and drives the system toward classically trivial configurations. In this work, we argue through a mix of analytical arguments and numerical simulations that this need not be the case in general, and propose a variation of the Bose-Hubbard sampling problem which resonantly couples the system to  highly lossy elements (in this case, harmonic oscillators in the form of superconducting cavities). Through a variety of numerical benchmarks we show that this open quantum system should also be extremely hard to simulate, and due to the expanded Hilbert space and need to average over many quantum trajectories for accurate results, we expect the system to become classically intractable at around two thirds the size of the equivalent unitarily evolving chain, and half the size of a comparable circuit of qubits enacting random discrete gates.

Since the lossy cavities are already included for state readout in any superconducting qubit implementation, the only additional experimental features required by our protocol are additional microwave signals to resonantly drive qubit-cavity interactions. Since these cavities are left idle throughout the evolution in other protocols, and are only populated for state measurement at the end of the evolution, in traditional unitary protocols fully half of the system's quantum degrees of freedom are left idle. In contrast, in our proposal they are integral to the system's dynamics, so our protocol thus nearly maximizes the quantum simulation complexity for a given hardware layout. Our results here are focused on superconducting qubit platforms due to the hardware efficiencies and relative ease of engineering complex quantum dynamics through dissipation \cite{kapit2017review}, but could easily be generalized to other quantum platforms such as trapped ions or neutral atoms. These results expand the space of interesting sampling problems, and suggest that a quantum advantage may be possible to demonstrate in smaller systems than previously thought.

This paper is organized as follows. We first describe our new protocol, then discuss important general considerations for sampling problems which include noise. We then simulate the dynamics of our protocol using experimentally realistic target parameters, and compute a series of key benchmark quantities to demonstrate classical hardness, including volume entanglement, signatures of quantum chaos in the form of distance from a Porter-Thomas distribution, number fluctuations, inverse participation ratio, heavy output generation and expected fidelity loss from various sources, both experimental and in simulation. Extrapolating from these, we provide estimates for expected classical simulation difficulty at larger system sizes, and show that, under the assumption that direct Hamiltonian time evolution is the most efficient simulation method, the system should become impossible to accurately simulate with near-term classical hardware for chains or grids of between 25 and 30 qubit-cavity pairs, depending on protocol details.

\section{Proposed protocol}

\begin{figure}
\includegraphics[width=3.25in]{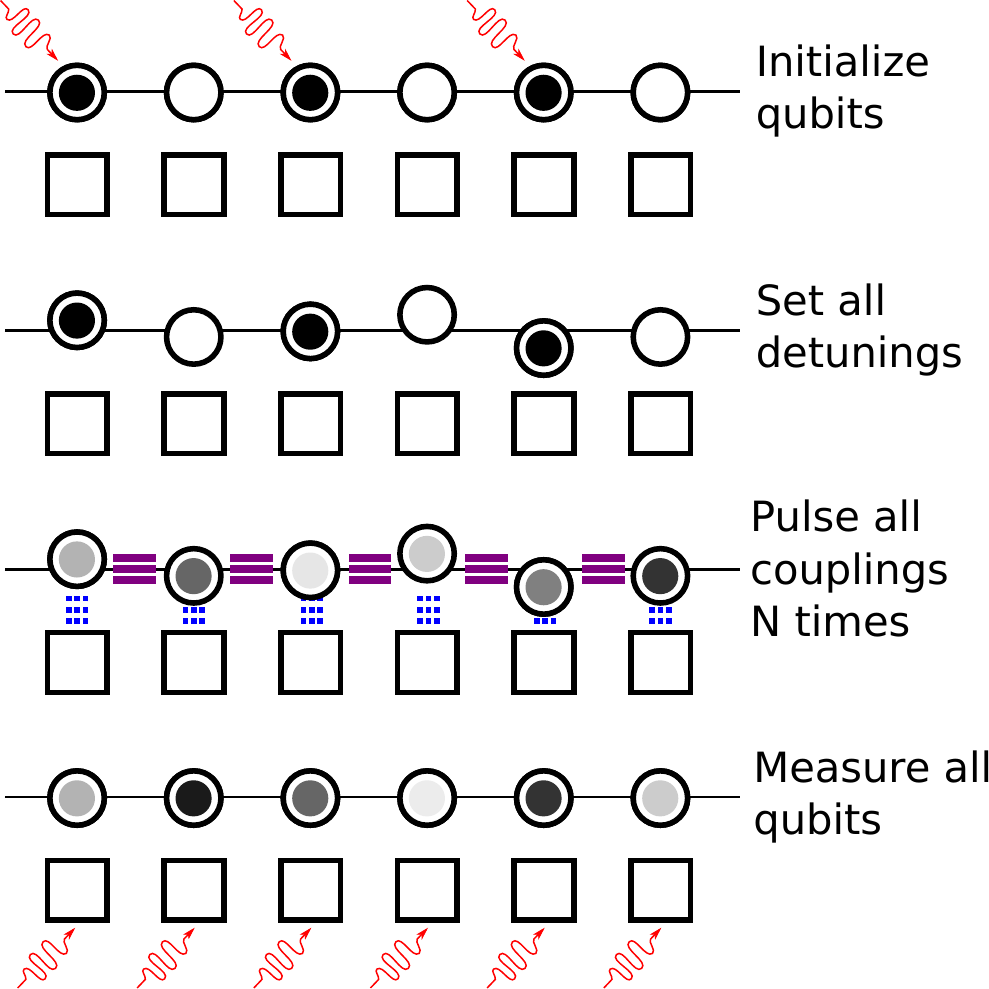}
\caption{Basic protocol studied in this work, an extension of the experiment reported in \cite{neillroushan2017}. As in the original work, a chain of qubits is initialized in a simple product state in the $z$ basis, a random set of detunings is applied to the qubits (circles), the nearest neighbor qubit-qubit exchange couplings (purple lines) are repeatedly pulsed on and off, and then the detunings are turned off and all qubits are measured in the $z$ basis. This program is repeated a sufficient times to estimate the fidelity with the aid of classical simulations. The key difference in our protocol is that driven sideband interactions (dashed lines), coupling the qubits to their readout cavities (boxes), are simultaneously turned on whenever the qubit-qubit couplers are, significantly changing the quantum dynamics and implementing a Hamiltonian where total photon number is no longer conserved. The magnitudes of all detunings and sideband interactions are weak compared to the qubit-qubit coupling terms, ensuring delocalized evolution and sharp resonance conditions in the qubit-cavity interactions.}\label{protfig}
\end{figure}

\begin{figure}
\includegraphics[width=3.25in]{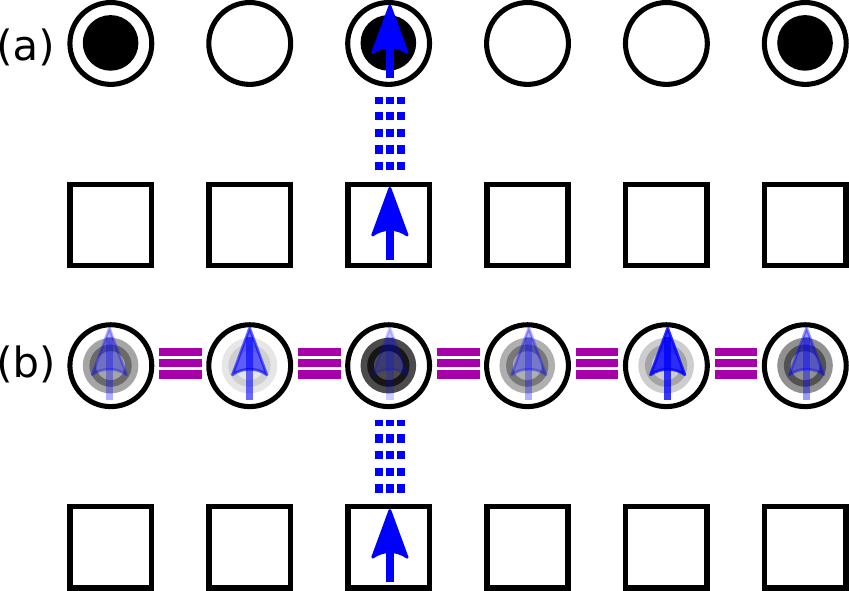}
\caption{Schematic picture of qubit-cavity interactions. In (a), the qubits (circles) are uncoupled from each other; as a result, the qubit-cavity drive (blue dashed lines) simply excites that qubit and its corresponding cavity (boxes), ignoring the state of the other qubits. A subsequent photon loss from the cavity thus acts as a local measurement of that qubit. In contrast, in (b) qubit-qubit exchange couplings (purple solid lines) are turned on at the same time as the qubit-cavity drive. In the limit that these couplings are much stronger than the qubit-cavity drive itself, the qubit-cavity drive can only couple to propagating modes within a narrow energy range, which have weight over the entire chain (represented by semi-transparent blue arrows at each qubit). Photon losses from the cavity then act as a measurement of a much more complex nonlocal operation, and do not necessarily disentangle the state. This property is vital for maximizing the simulation complexity of our noisy system, and more generally for employing noise to generate and stabilize nontrivial quantum states \cite{kapit2017review}.}\label{qcavfig}
\end{figure}

Quantum sampling problems based on unitary evolution amount to sampling from the distribution with probabilities $P_k$ of observing basis state $\ket{k}$ after evolving a known initial state with a potentially time-dependent $H \of{t}$ up to some time $T$. Sampling problems including noise are also based on sampling from the distribution $P_k$, which are in this case the diagonal entries of a density matrix evolving under the Lindblad equation \cite{gardinerzoller}:
\begin{eqnarray}\label{defL}
\partial_t \rho  = i \sqof{H \of{t}, \rho } + \sum_{i=1}^{K} \of{ O_i \rho  O_i^\dagger + \frac{1}{2} \cuof{O_i^\dagger O_i , \rho } }
\end{eqnarray}
Here, $K \propto L$ is the number of Lindblad operators and $L$ is the system size. For simplicity we assume that $H \of{t}$ can vary in time but that the Lindblad operators $O_i$ do not, though of course they may depend on time as well. Within this extremely general class of possible simulation problems, the protocol we consider in this work is a modification of the gmon chain experiment reported in \cite{neillroushan2017}. We begin with the $L$-qubit Hamiltonian
\begin{eqnarray}\label{HQ}
H_Q \of{ t} &=& - g \of{t} \sum_{i=1}^{L-1} \sqof{ a_i^\dagger a_{i+1} + {\rm H. c.} } \\
& & + \sum_{i=1}^L \sqof{h_i a_i^\dagger a_i - \sum_{n=2}^{n_{max}} \delta_n \ket{n_i} \bra{n_i} }. \nonumber
\end{eqnarray}
Here, $h_i$ are a set of local detunings, the $\delta_n$ are the qubit nonlinearities and $g \of{t}$ is a time dependent coupling strength which is ramped up and down, with the pulse waveform carefully optimized so that the population of $\ket{2}$ and $\ket{3}$ states is negligible at the end of each pulse (though the population of such states mid-pulse may be significant). In principle each qubit-qubit coupling can be tuned independently from the others, but we ramp them all up and down with the same profile for simplicity. Each qubit is weakly coupled to a lossy readout cavity; in the default protocol these terms do not appear in $H_Q$ because the cavities are only used for state measurement and do not effect the quantum evolution. We modify this protocol by including a set of driven qubit-cavity couplings, which couple each qubit to its lossy readout cavity via the Hamiltonian
\begin{eqnarray}\label{HQC}
H_{QC} \of{t} = \sum_{i=1}^L \sqof{ h_{Ci} a_{Ci}^\dagger a_{Ci} + \Delta a_{Ci}^\dagger a_{Ci} a_i^\dagger a_i } \\
+  \sum_{i=1}^L \sqof{ \Omega_{QCi}^{R} \of{t} a_{Ci}^\dagger a_i + \Omega_{QCi}^{B} \of{t} a_{Ci}^\dagger a_i^\dagger + {\rm H. c. }}. \nonumber 
\end{eqnarray}
Here the $h_{Ci}$ are a set of resonator detunings, $\Delta$ is the qubit-cavity dispersive shift and $\Omega_{QCi}^R$ and $\Omega_{QCi}^B$ are the amplitudes of the red and blue sideband qubit-cavity drives, respectively. These can be engineered \cite{murchvool2012,strandware2013,kapit2015,lima2018perfect} in the gmon architecture of flux tunable transmons qubits with fixed capacitive couplings to their cavities through oscillating the qubit energy near the difference of the qubit and cavity frequencies (red) or driving the qubit or cavity at frequencies near half the sum of the two frequencies (blue). For simplicity, we will consider only blue sideband protocols in this work (all $\Omega_{QCi}^R = 0$) since these terms are somewhat easier to engineer in a noise tolerant manner. Further, for reasons which will become clear below, we require that all couplings (qubit-qubit and qubit-cavity) are turned on \textit{simultaneously}, as sketched in FIG.~\ref{qcavfig}, rather than sequentially or in disconnected groups, as in gate model protocols. After being initialized in a simple product state (in the $z$ basis), the couplings are pulsed on and off for a total of $C$ cycles, at which point the states of all the qubits are measured in the $z$ basis. This sequence is repeated many times to generate an output sample, which is then compared to a theoretical model to calculate fidelity.

\section{General considerations for sampling problems with noise}

Before presenting the results of our numerical simulations, it is worth pausing to consider some of the important differences between noisy sampling problems and their purely unitary counterparts. In this section, we will discuss these differences, and argue a number of key points. First, we will demonstrate the perhaps obvious point that there exist nontrivial choices of the $\cuof{O_i}$ for which sampling the output distribution of (\ref{defL}) is at least as difficult as any unitary problem. Second, we will show that this is \textit{not} the case for some of the most natural choices, which include empirical models of random qubit error. Third, we will show that the worst cases of (\ref{defL}) are at most polynomially harder in total Hilbert space size than their unitary counterparts, and that realistic problems are likely to be more difficult by a factor which is polynomial in the size of the system and total evolution time. Following these results, we will outline key metrics for classical hardness that candidate protocols should satisfy, and then compute them explicitly in numerical simulations for the noisy sampling problem at the center of this work.

We begin by first noting that evolution under (\ref{defL}) for arbitrary $\cuof{O_i}$ has at least as much computational power as unitary quantum evolution, as shown by Verstraete \textit{et al} \cite{verstraete}, who provided an explicit construction for a set of Lindblad operators $\cuof{O_i}$ capable of universal quantum computation, even if the unitary Hamiltonian is zero. Thus, systems evolving under (\ref{defL}) can have at least as much computational power as unitary gate model quantum computation, and at minimum the worst cases of the noisy sampling problem should be extremely difficult to simulate on classical machines. Further, the operation of real, noisy quantum hardware is often well-approximated by (\ref{defL}), and topological error correction codes can be modeled through complex Lindblad operators; schemes to engineer self-correcting quantum codes \cite{fujiinegoro2014,brownloss2016} are examples of such an approach. These results further suggest that the general sampling problem of Lindblad evolution should be exponentially difficult for classical machines. Of course, simulating this evolution is \textit{not} exponentially difficult for a digital quantum computer \cite{wang2011quantum,di2015quantum,sweke2016digital}; by using ancilla qubits to model irreversible processes, one can accurately simulate dissipative evolution with polynomial overhead, at least for cases such as the one we consider here where the Lindblad operators are simple.

However, both these examples are obviously rather specialized, and in both cases the engineered Lindblad operators are irreducibly nonlocal. It is thus reasonable to ask how Lindblad operators deriving from a realistic noise model for modern quantum hardware will effect complexity, and in this limit things are naturally less clear cut. In many cases, the addition of noise simply makes the problem more trivial, and noisy elements which cannot create any type of correlations on their own are not good candidates for designing nontrivial sampling problems. For example, the addition of depolarizing noise (uncorrelated Pauli errors along $x$, $y$ and $z$ applied randomly at equal rates to each qubit) to random quantum circuits drives the system toward incoherent uniform randomness (IUR), a trivial distribution where all $P_k = 2^{-L}$ \cite{boixoisakov2016}. In fact, due to the chaotic nature of evolution in that system, to good approximation the final distribution is given by $\of{1 - P_{err} } \rho_{U} + P_{err} \rho_{IUR}$, where $P_{err}$ is the probability that at least one error has occurred in any of the qubits, $\rho_{IUR}$ is the incoherent random distribution and $\rho_U$ is the distribution which would result from noise-free evolution. We will show later in this work that realistic qubit error, in the form of white noise dephasing and photon loss, has similar effects on the evolution of an interacting Bose-Hubbard chain, with or without other nontrivial noise sources included in the sampling problem, though the fidelity loss from a single error depends on the type of error and may be somewhat less than 1. On general grounds, we would expect similar trivializing behavior from any set of Hermitian $\cuof{O_i}$ applied identically to all degrees of freedom in the system (since such operators create an incoherent random walk in Hilbert space), and the influence of many non-Hermitian $\cuof{O_i}$ choices applied identically everywhere should likewise drive the system toward trivial distributions.

That said, while these considerations pose serious challenges to crafting sampling problems where the noise is nontrivial, there is at least one key exception that offers reasons to be hopeful. Consider a quantum system simultaneously evolving under a continuously applied, delocalized many-body Hamiltonian $H$ (which may vary with time) and interacting with a bath that can be captured by a set of Lindblad operators $\cuof{O_i}$ which arise from local interactions between bath degrees of freedom and the constituent qubits. If these operators are simple Pauli matrices (potentially including non-Hermitian $\sigma_i^{\pm}$ terms), then we expect the resulting incoherent (though perhaps biased) random walk to simply push the system toward classically trivial states. Now imagine that the system is weakly coupled to the bath through local spin flips, resulting in transition rates which are sensitive to the energy difference between the given pair of states. If the system is delocalized, the resulting eigenstates are superpositions of many basis states (exponentially many for a general, delocalized many-particle system), and transitions between one eigenstate and another require operations to be performed across large fractions of the system, so for a transition induced by a local operator to be sensitive to energy changes in the system's state the operator must necessarily be modified into something extremely complex and nonlocal, with weight distributed across the system\footnote{The
``range" of these new operators depends on the details of the Hamiltonian and on the energy-dependence function which modulates the matrix elements (on general grounds, we expect slowly varying functions to correspond to shorter ranges than sharply peaked ones, based on the inverse polynomial splitting of propagating modes in the free particle case), but we will argue later in this work that it can be quite long, and thus, applications and measurements of these operators have highly nontrivial effects on the system's state and, being nonlocal, do not necessarily disentangle it.}.

The most natural example of such nonlocal operators arising from local couplings is a system's interaction with a low-but-finite temperature thermal bath, which has been shown to be extremely difficult to faithfully simulate \cite{jaschke2018thermalization}; while the thermal states of many-body systems can often be accurately simulated with quantum Monte Carlo if they lack a sign problem, the detailed time dynamics of thermalization beginning from an arbitrary initial condition cannot. And though these operators do not occur naturally in high-coherence quantum information platforms driven by oscillating fields, such as trapped ions or transmon qubits, they can be engineered straightforwardly by coupling the system to auxiliary, lossy elements (see \cite{kapit2017review} for a review), as illustrated in FIG.~\ref{qcavfig}. It is this type of system we choose to study, and we will show that configurations of this type are capable of generating complex quantum dynamics, even when the noise is strong and we expect multiple incoherent events to have occurred in the course of the evolution. 

Given that classically hard noisy protocols exist, and a likely route toward them via engineering effective global operations through resonant coupling to lossy subsystems, it is natural to ask how much more (or less) difficult simulation of these systems should be in comparison to unitary evolution. From the chaotic signatures presented below, we assume that the only classical algorithms to simulate the required sampling involve calculating the ideal probabilities $P_k$. Clearly, storing the full density matrix in Eq. (\ref{defL}) is horrendously inefficient (it has a memory cost proportional to $N_H^2$ for Hilbert space size $N_H$), since the protocol is designed to explore a large fraction of Hilbert space and thus $\rho$ will not be sparse. One can reduce the memory cost by using trajectory methods \cite{daley2014}. These schemes require only $O \of{N_H}$ in memory (since only a wavefunction needs to be stored), and evolving a single trajectory costs only $O \of{\of{T/dt} \times L \times N_H}$ in time (where $dt$ is a sufficiently small timestep), since each sparse matrix-vector multiplication requires $O \of{L \times N_H}$ operations. However, we have to sample a large number of trajectories $N_t$ to accurately solve (\ref{defL}). Let us assume we want to find all the $P_k$ over some restricted fraction of Hilbert space $A$, with dimension $N_{HA}$; in our case $A$ is the qubit subspace and $N_{HA} = 2^L$. As discussed in \cite{volokitinliniov2017} and other works, the \textit{worst case} estimate of $N_t$ is exponentially large, since we want $N_t$ to be large compared to the average per-trajectory variance $\delta P_k $ divided by $P_k$ itself, and $\avg{P_k} = 2^{-L}$. In this limit trajectory methods are hardly faster than density matrix evolution, though they do use substantially less memory.

However, the true scaling of the variance $\delta P_k$ is problem dependent and the worst case assumption may be exceedingly pessimistic. First note that to produce a sample we can output one bitstring with the correct probability from each trajectory. In addition, to produce a sample of size $M$ with fidelity $\alpha$, it suffices to sample $\alpha M$ bitstrings from the ideal distributions~\cite{villalonga2018flexible}. This upper bounds the number of quantum trajectories required. From a different point of view, for the delocalized system we consider in this work all $P_k \propto 1/ N_{HA}$ in a \textit{typical} trajectory, and therefore $\delta P_k \propto 1/N_{HA}$ and $N_t$ does not grow exponentially with system size (as shown below, we empirically find $N_t$ grows linearly or quadratically with the product of system size and evolution time, depending on the observable of interest). However, in cases where a typical trajectory has most $P_k$ values nearly equal to zero and a few values exponentially larger than $1/N_{HA}$ the variance may be larger, provided that the locations of the large $P_k$ values can vary substantially from one trajectory to the next. These arguments apply equally well to simulations based on matrix product states or similar constructions, as we describe toward the end of this work. We thus conclude that simulating noisy evolution at least as hard as noise-free Hamiltonian time evolution, and polynomially harder in the worst case.

Given all these considerations, we can wrap up our general discussion of noisy sampling problems with a set of benchmarks that must be met if we are to strongly believe that no polynomial classical algorithm could reproduce the output distribution. First, and most obviously, the evolving wavefunction should require an exponential amount of classical information to store. This requirement implies that the evolution should explore a large fraction of Hilbert space (as measured through inverse participation ratio \cite{canovirossini2011}), and achieve volume-law entanglement\footnote{In a 2d grid of locally coupled qubits, area-law entanglement, which is the maximum entanglement achievable for noisy evolution for sufficiently long times and large system sizes \cite{chan2018weak,li2018quantum,skinner2018measurement}, would also lead to a superpolynomially growing cost to store the wavefunction, scaling roughly as $e^{c \sqrt{N}}$ for some $c$. However, given that random qubit error reduces the fidelity by a factor which is exponential in the number of qubits (and not its square root), unless entanglement grows sufficiently quickly the fidelity of a simulation on real hardware could become vanishingly small by the time classical intractability is reached. It thus strikes us as sensible to require entanglement to scale with the volume in a 2d system as well.}, since states whose total entanglement does not grow exponentially with system size should in principle have an efficient classical representation, though actually finding such a representation in practice may be difficult. Second, the output distribution should be (informationally) easy to distinguish from classically trivial configurations, such as incoherent uniform randomness. It is desirable on general grounds if the evolving mixed state displays features of quantum chaos, such as a Porter-Thomas distribution of amplitudes and rapid scrambling of any initial information, since this strengthens expectations for classical simulation difficulty, but this is not a strict requirement; there are many quantum problems (such as finding the ground states of local Hamiltonians \cite{kempekitaev}) which are not necessarily chaotic but have no efficient classical solution.

Finally, it is worth pointing out one clear advantage of intentionally noisy evolution: the possibility of achieving nontrivial steady states, even when random qubit errors are taken into account. In a purely unitary protocol such as RQC or Bose-Hubbard evolution, introducing random qubit error in the form of losses or dephasing leads inevitably to a trivial final state at long enough times, typically either IUR or an entirely empty lattice. However, this is not the case if the random qubit noise is balanced by carefully tailored noise in auxiliary elements. As summarized by one of us in a recent review \cite{kapit2017review}, engineered dissipation can be an extraordinarily useful resource in quantum computing with superconducting circuits, and complex many-qubit states can be stabilized. Undoubtedly, variations of the protocols we explore here could lead to highly nontrivial long-time configurations. Finding such protocols is not our purpose here-- and indeed, the long time states of the protocols we do simulate are likely trivial-- but the possibility is worth keeping in mind for future work.

\section{Numerical results}

We now present the main results of this work: extensive numerical simulations of our protocol. Of necessity, the systems we consider-- linear chains with $L$ ranging from 4 to 11-- are relatively small, but since each site corresponds to a qubit-cavity pair, the system's total Hilbert space is much larger than for a qubit chain alone. We first describe our simulation methods and parameters in detail, then plot results for entanglement negativity, a collection of different statistical measures of the output distribution, and the expected fidelity loss (in comparison to the output of an ideal evolution) from various sources including approximations made in simulation and error processes in the quantum hardware itself.

\subsection{Simulation details}\label{simdetails}

We consider blue sideband protocols, initialized in simple product states of $L/2 - 1$ photons in the qubits (rounded down for odd $L$) with all cavities empty. In all cases we draw a random set of coupler pulses with $g_{max} = 2 \pi \times 40 {\rm MHz}$ and durations randomly chosen within the range from $20$ to $30$ ns; all couplers are identically ramped up and down using a symmetrized hyperbolic tangent profile. During each pulse the same set of qubit-cavity interactions are applied with $\Omega_{QC,max} = 2 \pi \times 3.0 {\rm MHz}$, with a slightly narrower ramp profile with the same duration. The qubit nonlinearity is $\delta = - 2 \pi \times 200 {\rm MHz}$, the qubit-cavity dispersive shift is $\Delta = 2 \pi \times 5 {\rm MHz}$. The cavity photon loss rate is chosen to be $\Gamma_C = 10 {\rm MHz}$. Where applicable, these parameters were all chosen to roughly match the experimental parameters used in the unitary protocol which this work builds upon \cite{neillroushan2017}; other parameters (such as the cavity loss rate and qubit-cavity interaction strengths) are chosen as ``typical" values for superconducting qubit experiments. We consider two variations of our protocol: Parametrization A, where all $h_i \in 2 \pi \times \cuof{-20, +20 } {\rm MHz}$ and all $h_{Ci} = 0$, and parametrization B, where all $h_i \in 2 \pi \times \cuof{-5, +5 } {\rm MHz}$ and where each $h_{Ci}$ is chosen to be equal to one of the $L$ eigenvalues of the single-particle hopping matrix with $g=g_{max}$ (these assignments are randomized from one protocol instance to the next). Qubit and cavity detunings are fixed through all $N_c$ cycles of evolution.

We track various observables over 12 full cycles of evolution. For context, we note that assuming $g_{max} = 2 \pi \times 40 {\rm MHz}$ and $20 {\rm ns} \leq t_{cycle} \leq 30 {\rm ns}$, between $L/4$ and $L/3$ cycles are likely sufficient to fully entangle an $L$-site chain, as observed indirectly in \cite{neillroushan2017}. Given that $g \of{t}$ is ramped up and down over the course of a cycle, an average cycle time of 25 ns roughly corresponds to between 4 and 5 times $\avg{g}^{-1}$, a relatively long evolution time. A full 12 cycles thus amounts to an average of around 50 $\avg{g}^{-1}$, and three times the cavity photon lifetime.

To simulate the dynamics of our protocol, we use an event-driven quantum trajectory method as outlined in \cite{daley2014} to integrate the Lindblad equation beginning from a simple product state at $t=0$. To simplify the calculation, we make two  approximations. First, we truncate the cavity Hilbert space to include at most one photon per cavity, and cap the maximum number of cavity photons at a fixed value, respecting the fact that the cavities are lossy, begin in an empty state, and are coupled relatively weakly to the qubits, so their average photon populations should be low. We repeat our calculations with varying maximum cavity photon number, and track the fidelity loss from the truncation as a way of estimating the likely number of photons that would need to be kept in simulations at larger $L$.

Our second approximation is to truncate the qubit Hilbert space to zero or one photon per qubit, and in doing so we include additional qubit-qubit interactions (computed in second order perturbation theory) to account for our having integrated out states $\ket{2}$ and higher. As described in \cite{neillroushan2017} this is not expected to be a quantitatively good approximation for long times or large $L$, but it should not qualitatively change the behavior we are primarily interested in, such as bipartite entanglement, information scrambling, inverse participation ratios, and so on. We make this approximation primarily to avoid having to perform the complex task of pulse shaping to suppress local $\ket{2}$ and $\ket{3}$ states, which would be a substantial effort ultimately not relevant to the conclusions we make in this work. Specifically, perturbatively eliminating the $\ket{2}$ state generates nearest neighbor potential interactions and a mediated hopping term. For a given three sites these terms take the form
\begin{eqnarray}\label{Hpert}
H\of{t} &=& - g \of{t} \of{a_1^\dagger a_2 + a_2^\dagger a_3 + {\rm H. c.} } - \delta \sum_{i=1}^3 \ket{2_i} \bra{2_i} \\
&\to& - g \of{t} \of{\sigma_1^+ \sigma_2^- + \sigma_2^+ \sigma_3^- + {\rm H. c.} } \nonumber \\
& & + \frac{4 g \of{t}^2}{\delta} \sqof{\of{n_1 n_2 + n_2 n_3} + \frac{1}{2} \of{\sigma_1^+ n_2 \sigma_3^-  + {\rm H. c.}}  }, \nonumber
\end{eqnarray}
where $n_i \equiv \of{\sigma_i^z + 1}/2$. Our total Hamiltonian in simulation is thus equal to:
\begin{eqnarray}\label{Htotal}
H \of{t} &=& \sum_{i=1}^{L-1} \of{- g \of{t} \of{\sigma_i^+ \sigma_{i+1}^- + {\rm H. c.}} + \frac{4 g \of{t}^2}{\delta} n_i n_{i+1} } \nonumber \\
& &+  \frac{2 g \of{t}^2}{\delta} \sum_{i=1}^{L-2} \of{\sigma_i^+ n_{i+1} \sigma_{i+2}^-   + {\rm H. c.}}   \\
& &+ \sum_{i=1}^{L} \sqof{h_{i} n_{i} + h_{Ci} n_{Ci} + \Delta n_{i} n_{Ci}  } \nonumber \\
& &+ \Omega \of{t} \sum_{i=1}^L \of{\sigma_{i}^+ \sigma_{Ci}^+ + {\rm H. c.}}. \nonumber
\end{eqnarray}
It is this time-dependent Hamiltonian, extended to larger chains, that we use in our simulations. Note that the Hamiltonian is not symmetric about half filling ($L/2$ photons in the qubits), as is to be expected from the underlying Bose-Hubbard model it approximates. Further, for the parameters we choose, the interaction terms in the first and second lines are not small, for while they are smaller than $g \of{t}$ they are larger than the disorder strength and qubit-cavity interactions, and thus play a significant role in the physics. However, even in the limit of $\delta \to \infty$ where the interactions vanish and the isolated chain is integrable, interactions with the cavities break integrability and would likely still lead to the quasi-chaotic dynamics we observe here.

Beyond these approximations we use standard methods to simulate the system's dynamics, integrating (\ref{defL}) using 4th-order Runge-Kutta methods beginning from a simple product state with a fixed number of photons in the qubits and all cavities empty. The system's full density matrix is computed by averaging the sum of $\ket{\psi \of{t}} \bra{\psi \of{t}}$ over many randomized trajectories, which is then used to compute expectation values, entanglement measures, and so forth. 

\subsection{Negativity}

\begin{figure*}
\includegraphics[width=3.0in]{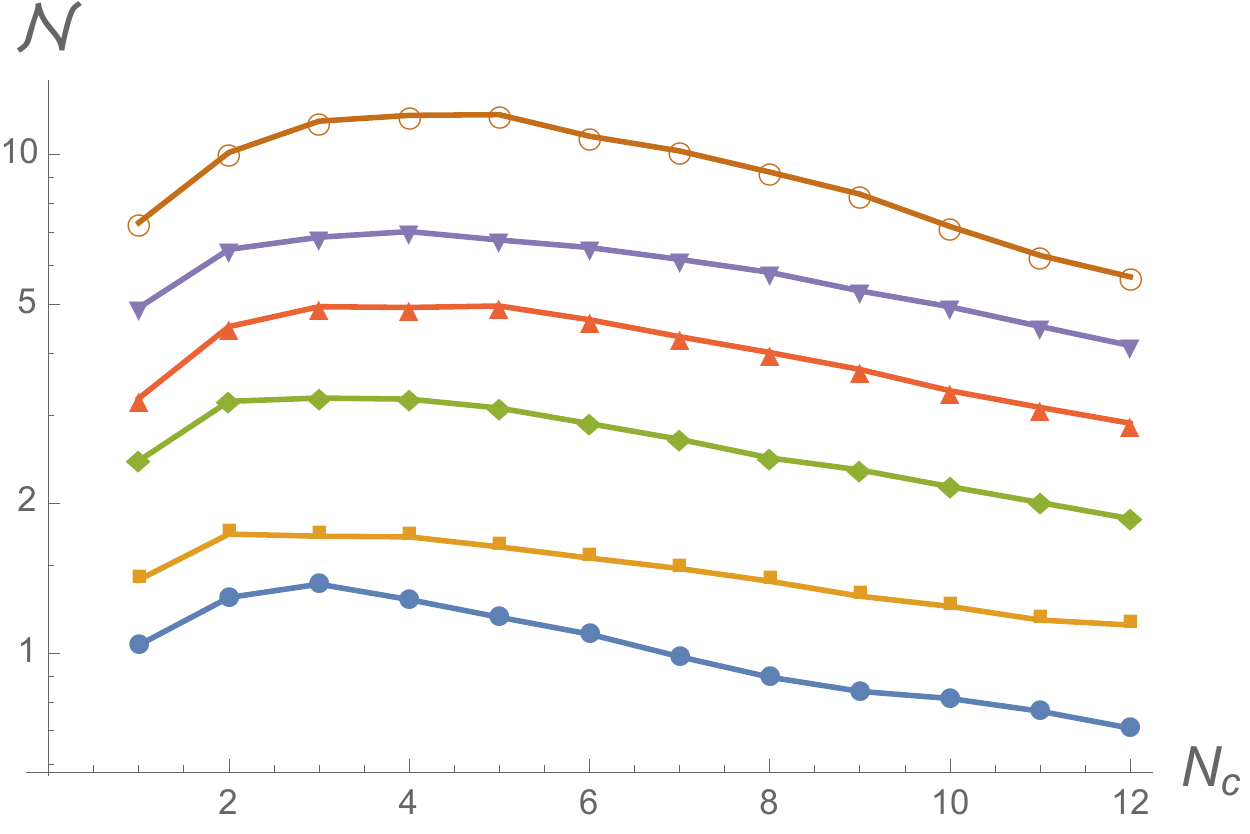}
\includegraphics[width=3.0in]{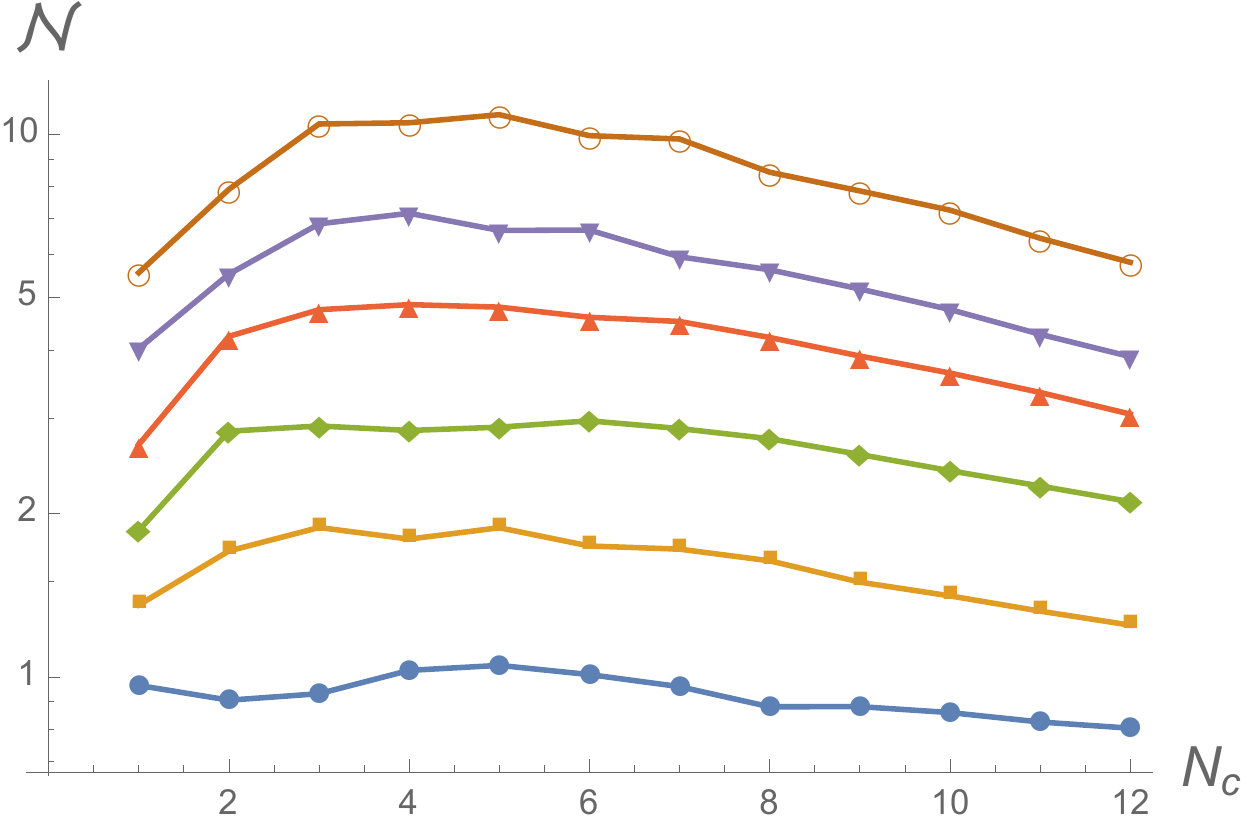}
\vspace{0.1in}
\includegraphics[width=3.0in]{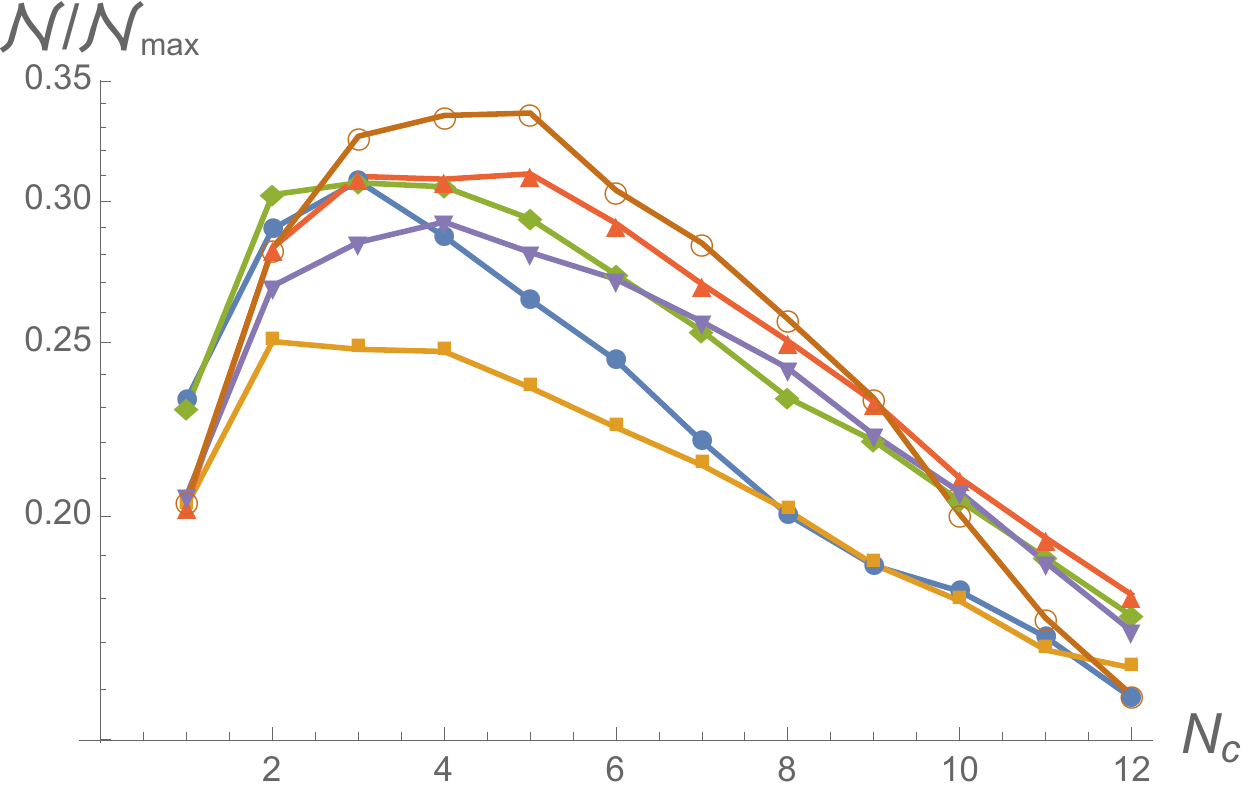}
\includegraphics[width=3.0in]{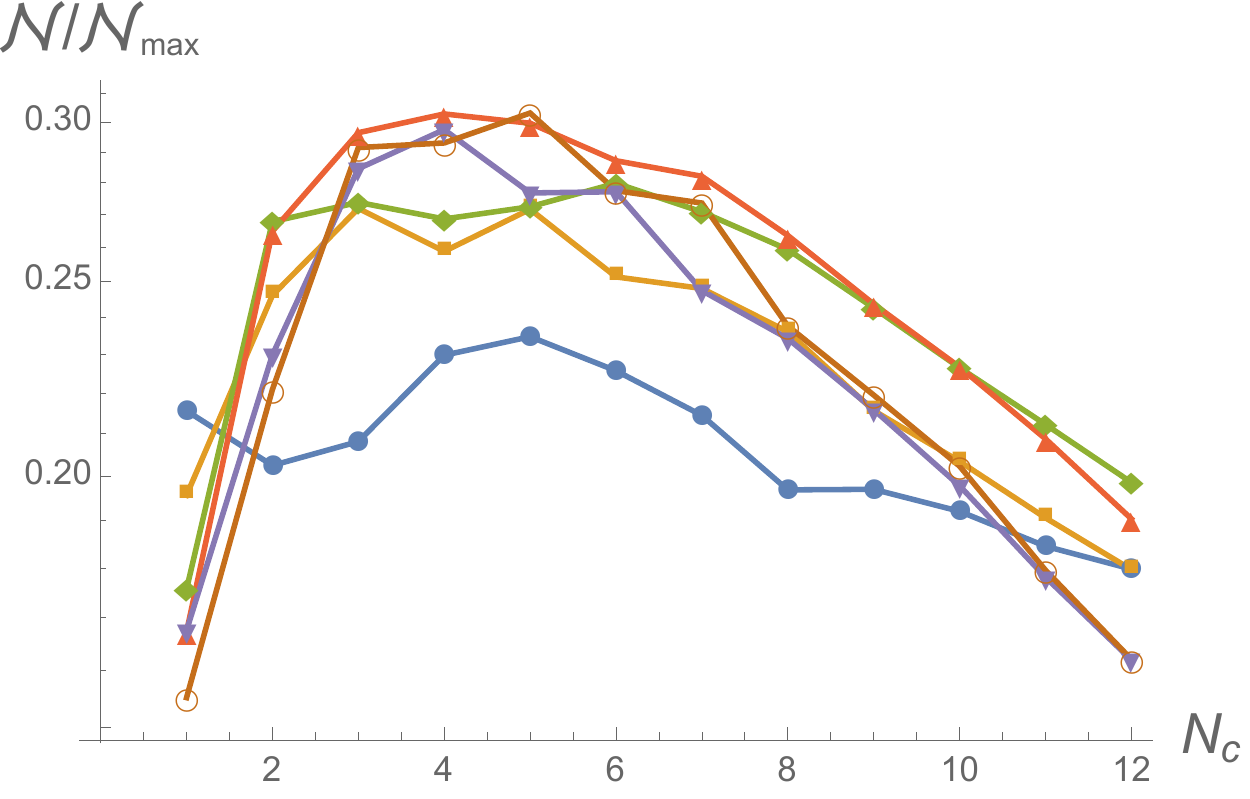}
\caption{Full system entanglement negativities after $N_c$ cycles of evolution. In the top row, we plot $\mathcal{N}$ for parametrizations A (left) and B (right; see the ``Simulation details" subsection for specific value ranges), and in the bottom row we plot the same quantity divided by the maximum possible negativity $\mathcal{N}_{max} = \of{1/2} \of{\sqrt{N_{H,eff}}- 1}$, where we used $N_{H,eff} = \of{1 + L} \times 2^L$ as the cavity photon population is kept low by photon loss and the full cavity Hilbert space is not explored. In this and all subsequent figures unless otherwise noted, $L=4$ is plotted with blue filled circles, $L=5$ gold boxes, $L=6$ green diamonds, $L=7$ red triangles, $L=8$ purple triangles, and $L=9$ brown open circles. The results in this and all subsequent figures are averaged over many random protocol instances. Aside from an even-odd effect where odd $L$ negativities tend to be larger, $\mathcal{N}/\mathcal{N}_{max}$ remains approximately constant as $L$ increases, showing that the system achieves volume entanglement at intermediate times, though entanglement does begin to decay after a handful of cycles due to continuous photon loss from the cavities. While we observe no saturation of $\mathcal{N}$ with increasing $L$, this should occur at some sufficiently large $L_{max}$ (see discussion in text), though we expect $L_{max}$ to be large enough that classically simulating the system's evolution will be impossible on any near-term classical computer.}\label{negfig}
\end{figure*}

The first, and arguably most important, quantity we measure is entanglement, since one can usually find efficient classical representations for weakly entangled states. To measure entanglement in our system, we use the bipartite negativity \cite{vidalwerner2002,eltschkasiewert2013}, $\mathcal{N} \equiv \of{1/2} \of{ \abs{\abs{\rho^{T_{A}} } }- 1}$ where $\rho^{T_{A}}$ is the partial transpose of the density matrix $\rho$ relative to subsystem $A$ and $\abs{\abs{\rho^{T_{A}} } }$ is the sum of the absolute values of its eigenvalues. The negativity, while expensive to compute (since it requires fully diagonalizing the density matrix of the full system), is equally well-defined for pure and mixed states; the more commonly used Von Neumann and Renyi entropies only measure entanglement accurately for pure states. A nonzero negativity is a sufficient, if not necessary, condition for quantum entanglement. For a perfect bipartition of the system the negativity is bounded by $\mathcal{N}_{max} = \of{1/2} \of{\sqrt{N_H}- 1}$, where $N_H$ is the Hilbert space size of the full system. If the system's negativity grows exponentially with $L$, then it obeys volume-law entanglement and it is extremely unlikely that any efficient classical representation exists for its state.

In FIG.~\ref{negfig}, we plot the bipartitie negativity $\mathcal{N}$ and the ratio $\mathcal{N}/\mathcal{N}_{max}$, where $\mathcal{N}_{max}$ is computed with $N_H = \of{1 + L} \times 2^L$, since we assume the resonator population is low. To keep the Hilbert space sizes approximately equal the system is partitioned such that partition $A$ contains all of the cavities and $\of{L - 3}/2$ qubits (fractions rounded up), with the remaining qubits placed in partition $B$; we make this choice because our nonlocal constraint on the maximum number of cavity photons makes it impossible to partition the cavity Hilbert space efficiently. As shown in the figure, the system rapidly achieves volume-law entanglement, and at least within the computationally accessible range of $L \leq 9$, even-odd effects aside there are no obvious trends in the scaling which suggest entanglement is beginning to saturate as $L$ increases. Our studies of entanglement are limited to $L=9$ and below due to the exploding cost of storing the full density matrix, which, assuming a maximum of 2 photons in the cavities, is almost 9 GB for $L=9$ and a bit over 52 GB for $L=10$. 

\begin{figure*}
\includegraphics[width=3.0in]{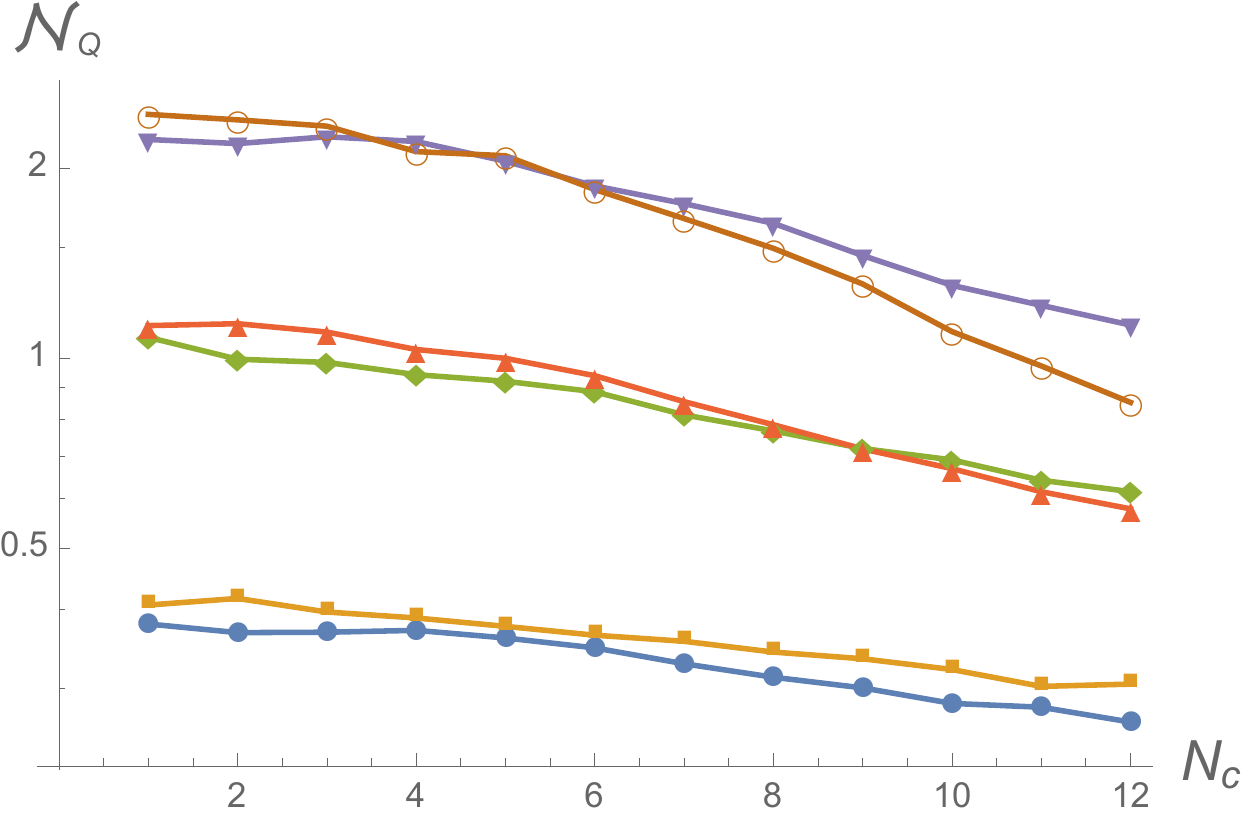}
\includegraphics[width=3.0in]{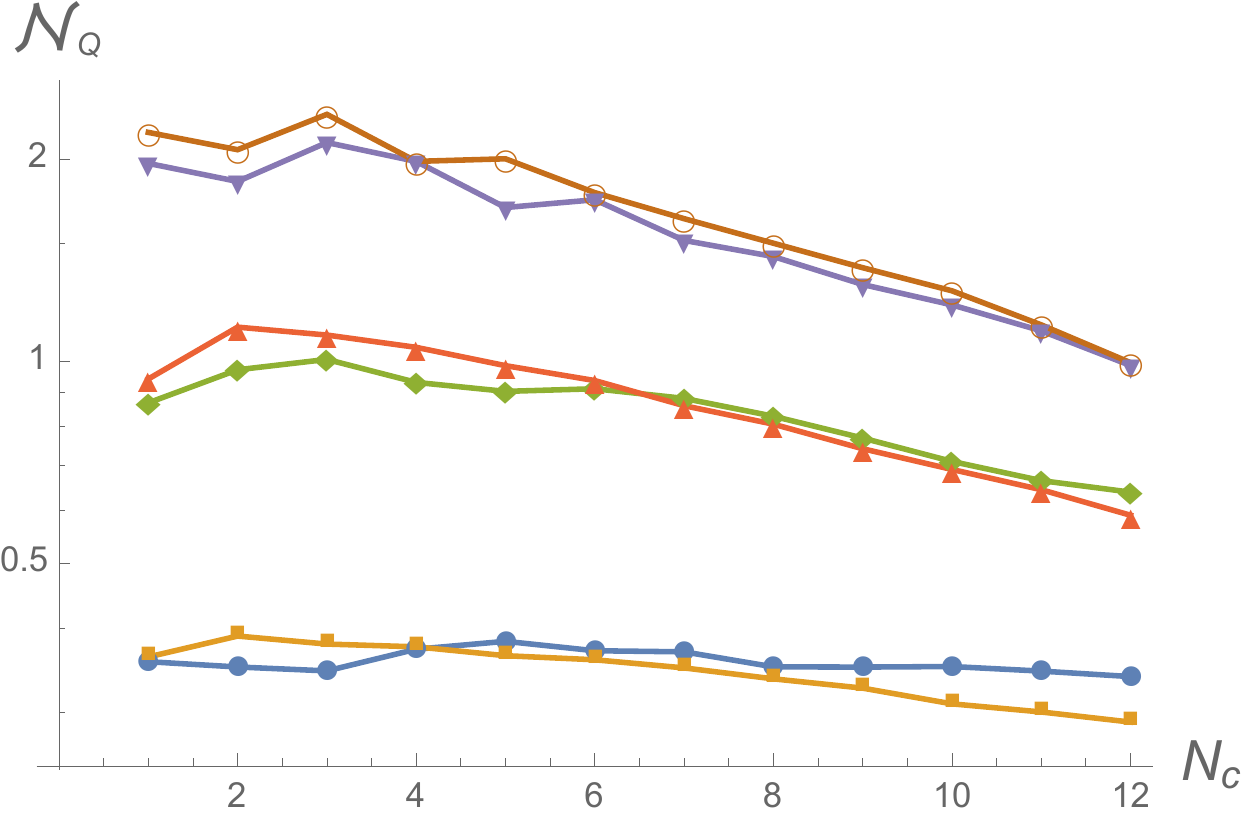}
\vspace{0.1in}
\includegraphics[width=3.0in]{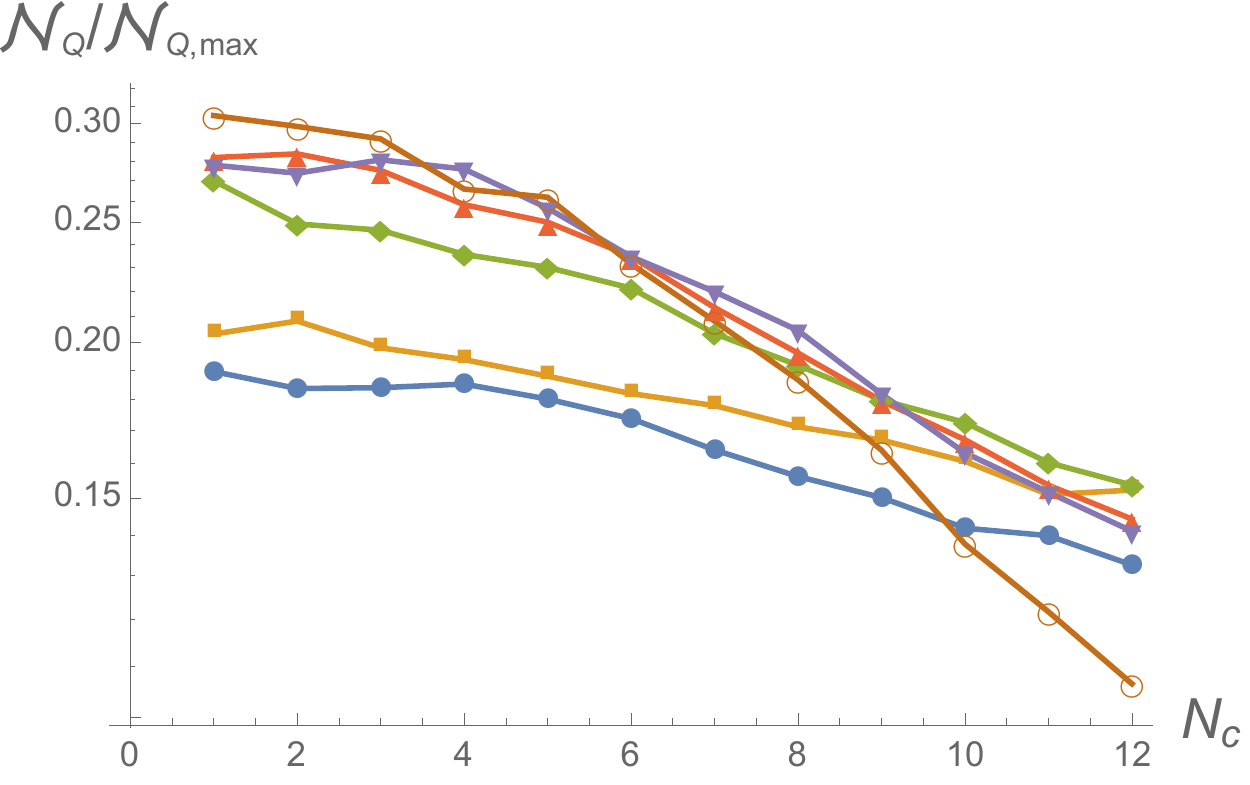}
\includegraphics[width=3.0in]{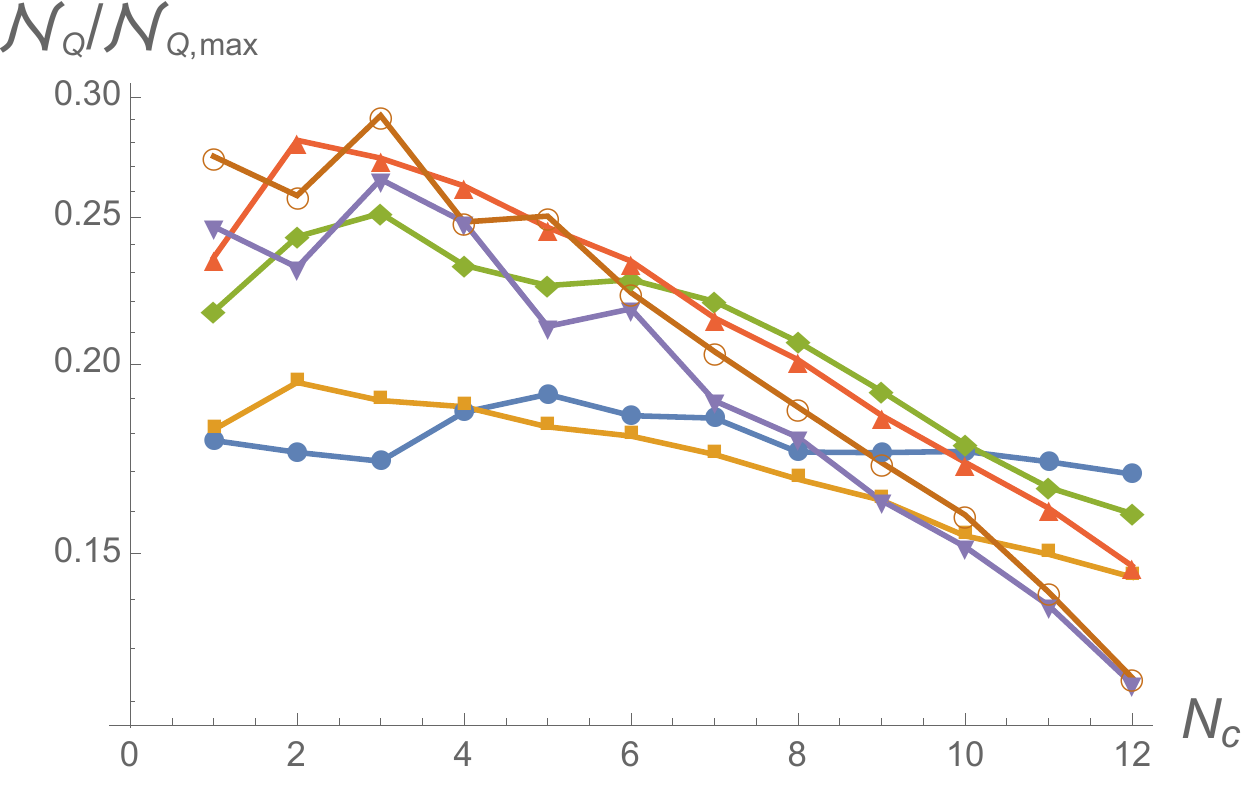}
\caption{Qubit subsystem entanglement negativities after $N_c$ cycles of evolution. In the top row, we plot $\mathcal{N}_Q$ for for parametrizations A (left) and B (right; see the ``Simulation details" subsection for specific value ranges), and in the bottom row we plot the same quantity divided by the maximum possible subsystem negativity $\mathcal{N}_{Q,max} = 2^{ {\rm floor} \of{L/2} -1}$. For the system sizes studied $\mathcal{N}_Q$ tends to continuously decay with increasing $N_c$ due to interaction with the cavities, as tracing out the cavities to calculate $\mathcal{N}_Q$ acts as a measurement on the qubit subsystem (albeit a complex, nonlocal one) even if no photon loss has occurred. However, it still clearly grows as a volume law, indicating the quantum state of the qubits remains extremely complex even with the cavities traced out. Note that methods for simulating time evolution which scale exponentially in bipartite entanglement, such as matrix product state representations \cite{orus2014practical}, will scale exponentially in the full system $\mathcal{N}$ and not merely the qubit subsystem entanglement plotted here.}\label{negQfig}
\end{figure*}

We can further probe the entanglement generated in our system by tracing out the cavities before computing $\mathcal{N}$, leaving a reduced negativity $\mathcal{N}_Q$ which captures the entanglement between two halves of the qubit subsystem. While this is not a useful metric for predicting the ultimate classical simulation difficulty in an MPS or PEPS-type simulation scheme (where the difficulty scales with the total bipartite negativity, not just the qubit subsystem's contribution), showing volume-law scaling of $\mathcal{N}_Q$ further bolsters our argument above that photon loss in the cavities does not fully disentangle the state. Note also that, since tracing out the cavities is equivalent to making measurements on the state (though the effect of these measurements is nonlocal as described above), we expect $\mathcal{N}_Q$ to be smaller in this calculation than it would be for an isolated, unitarily evolving chain, even before any photon losses have not occurred. In FIG.~\ref{negQfig} we show the results of this calculation. The observed subsystem negativities at intermediate times (eight cycles) are an average of nearly three times smaller than those computed for the purely unitary chain (where there are no measurement effects), but still grow exponentially with $L$, demonstrating that the quantum state of the qubits is extremely complex.

\subsection{Large-$L$ limits on entanglement}

A natural objection to this proposal is that continuous photon loss from the cavities will ultimately limit entanglement growth in the chain once $L$ becomes sufficiently large \cite{poulin2010lieb,barthel2012quasilocality,kliesch2014lieb,aolita2015open}. This in turn calls the ultimate difficulty of the problem into question, since states with bounded entanglement often have efficient classical representations through matrix product states or similar constructions \cite{orus2014practical}. Further, recent studies in random quantum circuits have shown that continuous (deterministically applied) measurement limits entanglement growth to an area-law \cite{chan2018weak,li2018quantum,skinner2018measurement}, a potentially trivializing effect if entanglement were to saturate at a small enough $L$ within reach of classical machines. Rigorously determining this limit for our protocol given realistic circuit parameters is an exceptionally difficult problem we will not attempt to answer, so instead we will consider two methods for roughly estimating it, and show that both arguments suggest that this $L$ can easily pushed into ranges beyond the simulation capacity of any forseeable classical computer.

Inspired by the lower bound calculated in \cite{zhang2018information}, we can provide a lower bound for the maximum length scale for correlations as follows. Let us imagine the Lieb-Robinson velocity for information propagation is $v$, photon losses occur at an average rate $\avg{n_{cav} } \Gamma_C$, where $\avg{n_{cav}}$ is the average photon density in a cavity during the evolution. Let us further assume a single loss is sufficient to fully scramble the state, as it does in RQC. Then the maximum length $L_{max}$ is given by the distance information can propagate before a single loss has occurred anywhere in the system; since these losses occur at a total rate $L \avg{n_{cav} } \Gamma_C$, and the time to entangle one end of the chain with the other is $t= L/ v$, we find $L_{max} \simeq \sqrt{ v / \avg{n_{cav}} \Gamma_C }$. For the gmon chain, $v$ can be estimated from the inverse of the time per iSWAP operation induced by the qubit-qubit couplers, which is around 3.5 ns assuming $g_{max} = 2 \pi \times 40 {\rm MHz}$, a ramp profile similar to that used in \cite{neillroushan2017}, and that all couplers are turned on simultaneously. $\avg{n_{cav}}$ is highly protocol dependent but a decent rough estimate is 0.05-0.1 based on the results detailed below, and $\Gamma_C = 10 {\rm MHz}$ is a typical loss rate in a readout cavity. This places $L_{max} \sim 17-24$; as shown toward the end of this work, the upper end of that scale may push into the limits of what is possible to simulate on near-term classical supercomputers. Further, if our expectation that the classical simulation difficulty scales exponentially in $L_{max}$ is correct, fairly small reductions in $\Gamma_C$ can increase the difficulty enormously.

\subsection{Negativity after a single photon loss}

However, the assumption that a single loss disentangles the state is empirically false for our protocol, so $L_{max}$ could be much larger. A plausible reason for this, introduced earlier in the general considerations section, is illustrated in FIG.~\ref{qcavfig}. Let us for the moment ignore interactions and disorder, and imagine the photons in the chain to be non-interacting bosons. Let us further assume, as mentioned above, that all terms are operated simultaneously, and the waveforms $\Omega_{QCi} \of{t}$ are shaped such that they are only nonzero when $g \of{t}$ is nonzero. During the evolution, if we assume $\Omega_{QC} \of{t}$ and $\Gamma_C$ are weak compared to $g_{max}$, then for a given qubit-cavity interaction we only have a significant probability of adding or removing a photon from the chain (and adding one to the cavity) if the total energy change in system is smaller than the minimum of $\Omega_{QC}$ and $\Gamma_C$. However, since the system is delocalized this condition can only be satisfied if the photon is added to or removed from a propagating mode, which has approximately equal weight over the entire lattice. A subsequent loss from the cavity, in other words, thus measures a highly nonlocal operator, and such measurements need not disentangle the state. The maximum length scale in this limit should be set by the mode splitting, which is approximately $5.8 g_{max} / L$ near the center of the band for a 1d chain. Requiring that the loss rate is less than half this gives $L_{max} \simeq 72$ from the parameters listed above, a much higher estimate than the lower bound of the previous paragraph.

Of course, interactions, disorder and the qubit-cavity dispersive shift all complicate this estimate, and the true value of $L_{max}$ probably lies somewhere in between the two predictions. Nonetheless, it is clear from these arguments that $L_{max}$ can be increased by reducing $\Gamma_C$, and assuming exponential difficulty scaling such reductions could push $L_{max}$ into a classically intractable range fairly easily. Furthermore, all of these concerns are moot in a 2d implementation, where a grid of $5\times 5$ or $4 \times 7$ qubit-cavity pairs will likely be sufficient to reach classical intractability (see the classical difficulty estimates section near the end of this work for the origins of this estimate) without any worries about the maximum range of correlations. Thus, while the evolution in our noisy protocol ultimately saturates in finite-ranged correlations, the range of those correlations can be quite long, and this effect does not keep this protocol from being a good candidate for simulating a classically hard quantum sampling problem with real quantum hardware.

\begin{figure}
\includegraphics[width=3.0in]{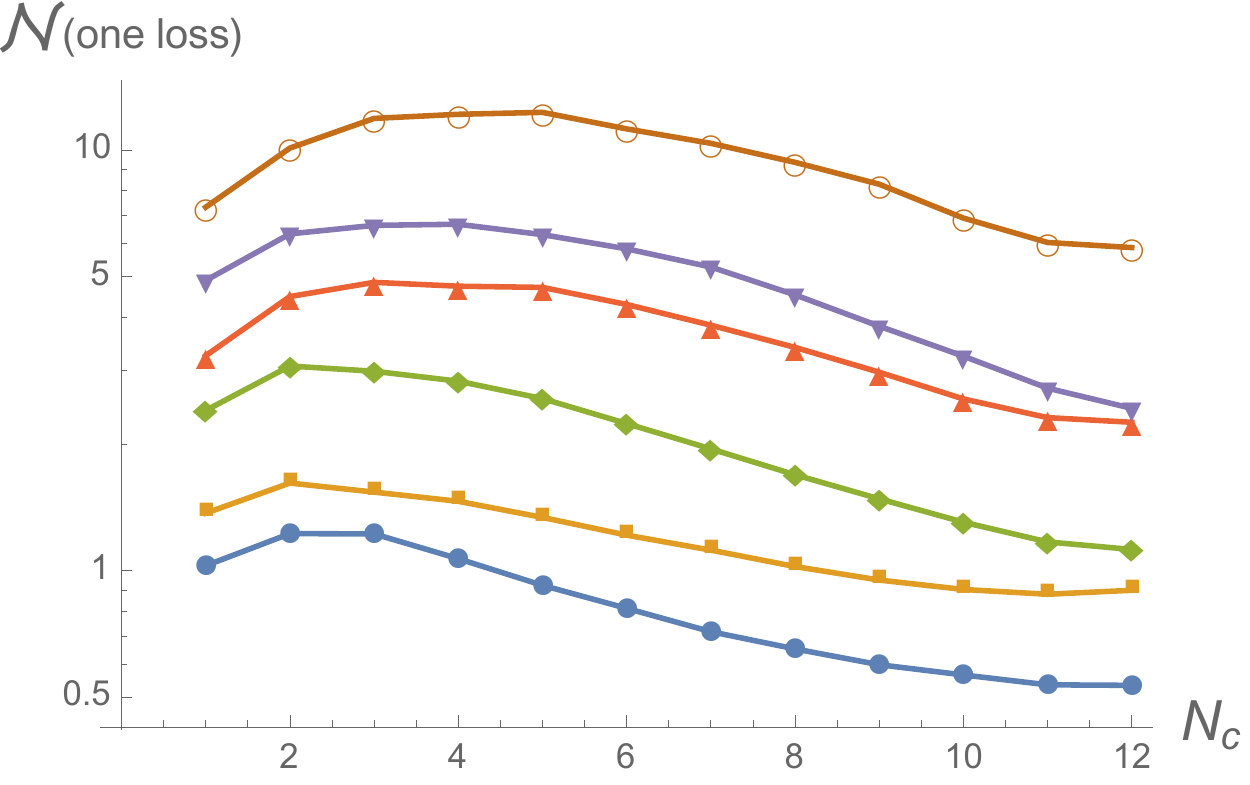}
\includegraphics[width=3.0in]{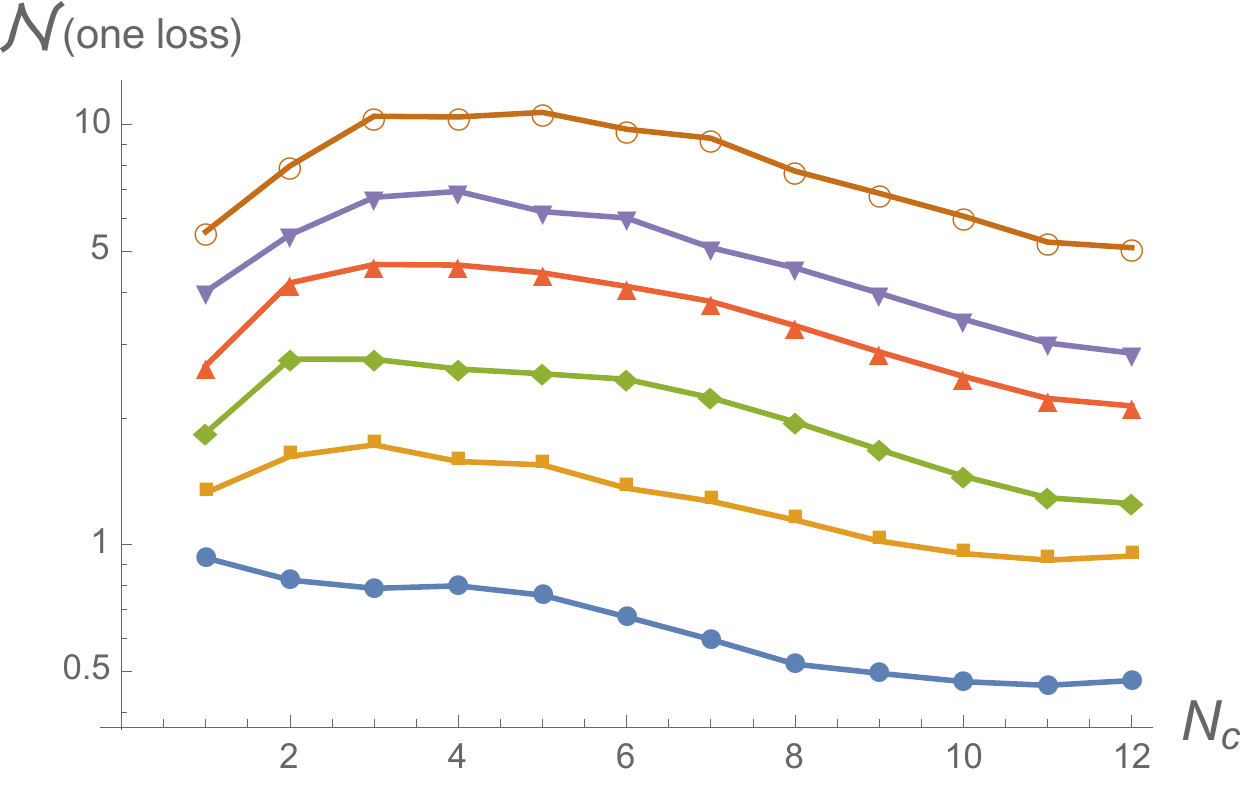}
\caption{Entanglement negativities averaged over only those trajectories where a single cavity photon loss has occurred during the 12 cycles, for parametrizations A (top) and B (bottom). These results show that substantial entanglement persists even after a photon loss (which acts as a local measurement of a cavity, but an effective nonlocal measurement of the qubits) has occurred, and that a single incoherent event does not decorrelate the state.}\label{onelossnegfig}
\end{figure}

To support this prediction, we take advantage of the fact that quantum trajectory simulations allow us to precisely track the number of photon losses, and present the entanglement negativity calculated from an average of only those trajectories where precisely one photon has been lost by the end of 12 cycles. To create this ensemble we generate a large number of trajectories using the same method as in the full simulation, but only include those where one photon has been lost in the subsequent averaging to construct the density matrix $\rho$. We plot the results of these calculations in FIG.~\ref{onelossnegfig}; as seen in the figure the system appears to maintain volume entanglement even after a photon loss has occurred, and in fact the final entanglement at 12 cycles is slightly \textit{larger} than in the full simulation for large $L$, which we assume reflects the fact that an average of more than one loss has occurred by that point in the full simulation. These results indicate that, unlike RQC, while cavity photon loss in our system does reduce entanglement, it does not completely destroy it, nor does it decorrelate the state. This suggests that the lower bound on the maximum range of correlations $L_{max}$ calculated in the previous subsection is too low, and that volume-law entanglement should persist in this system to much longer chains, likely beyond the scope of classical simulation.

\subsection{Output distribution: number fluctuations, distance from Porter-Thomas and incoherent uniform randomness}





\begin{figure*}
\includegraphics[width=2.25in]{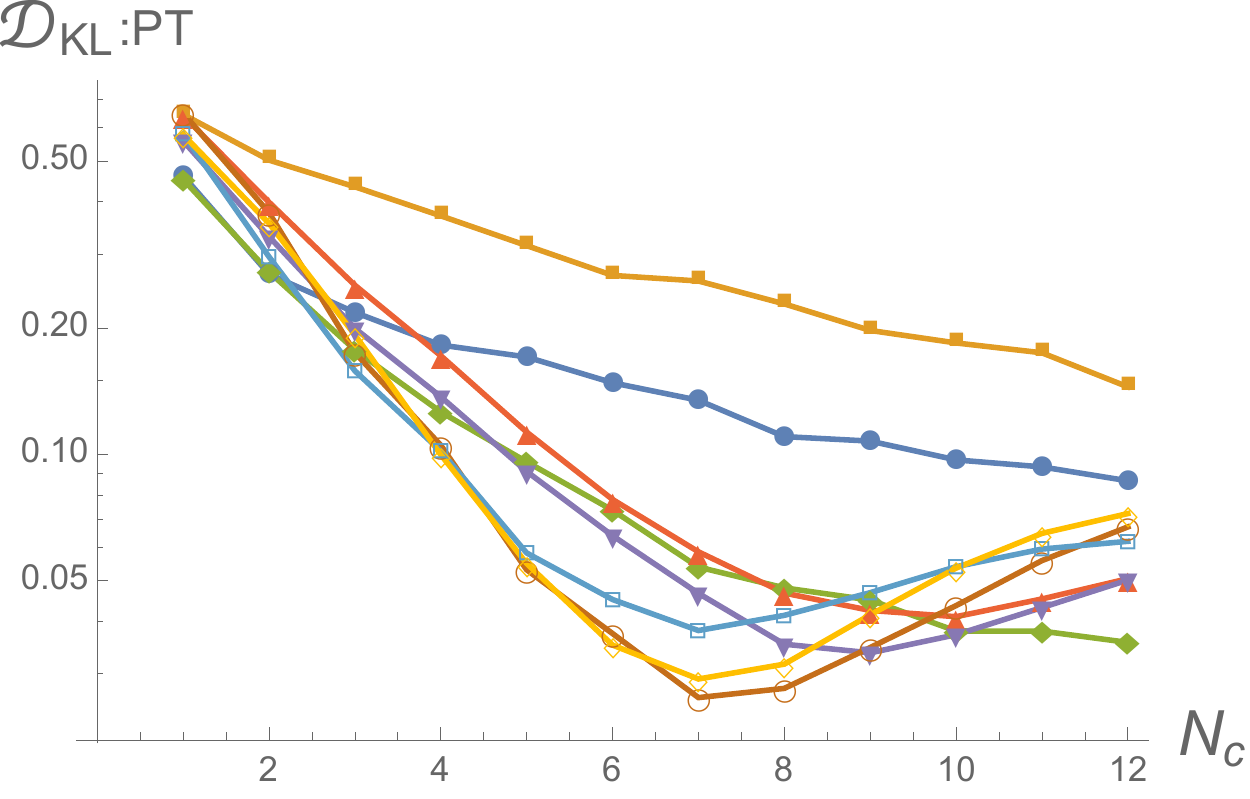}
\includegraphics[width=2.25in]{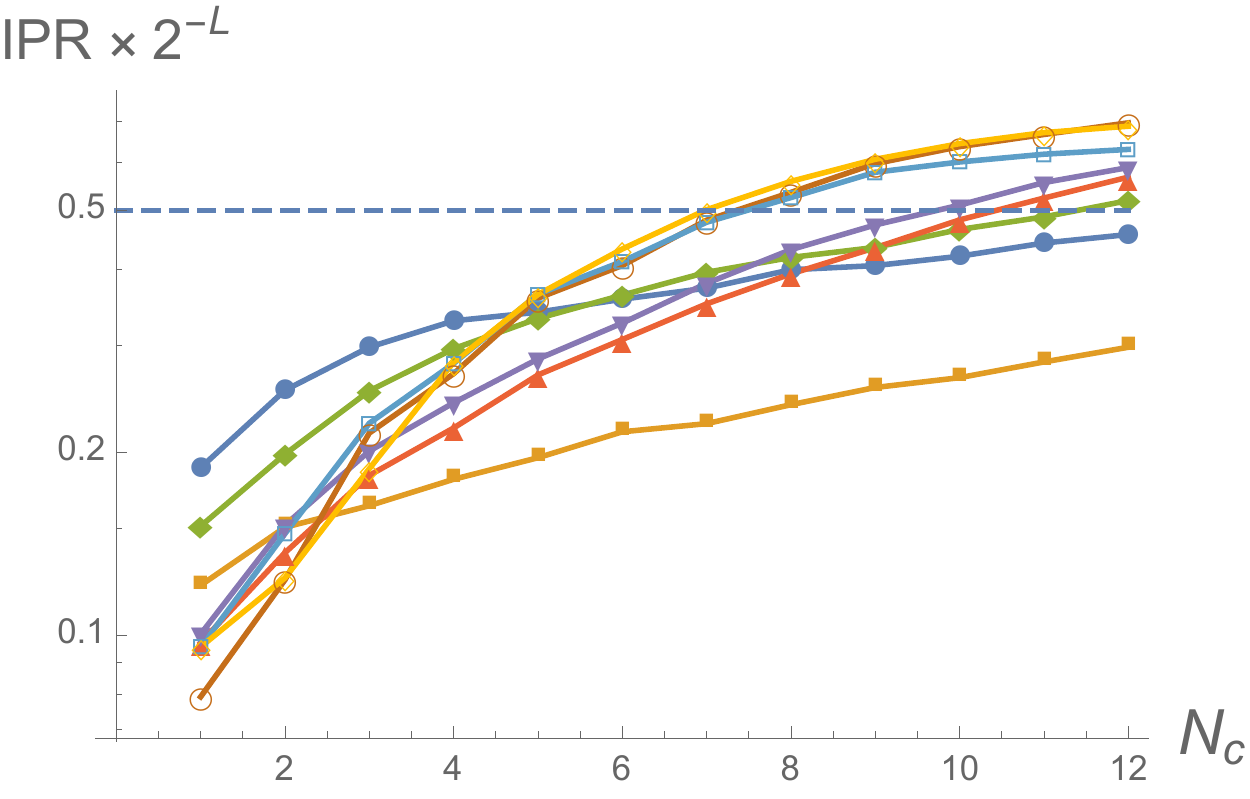}
\includegraphics[width=2.25in]{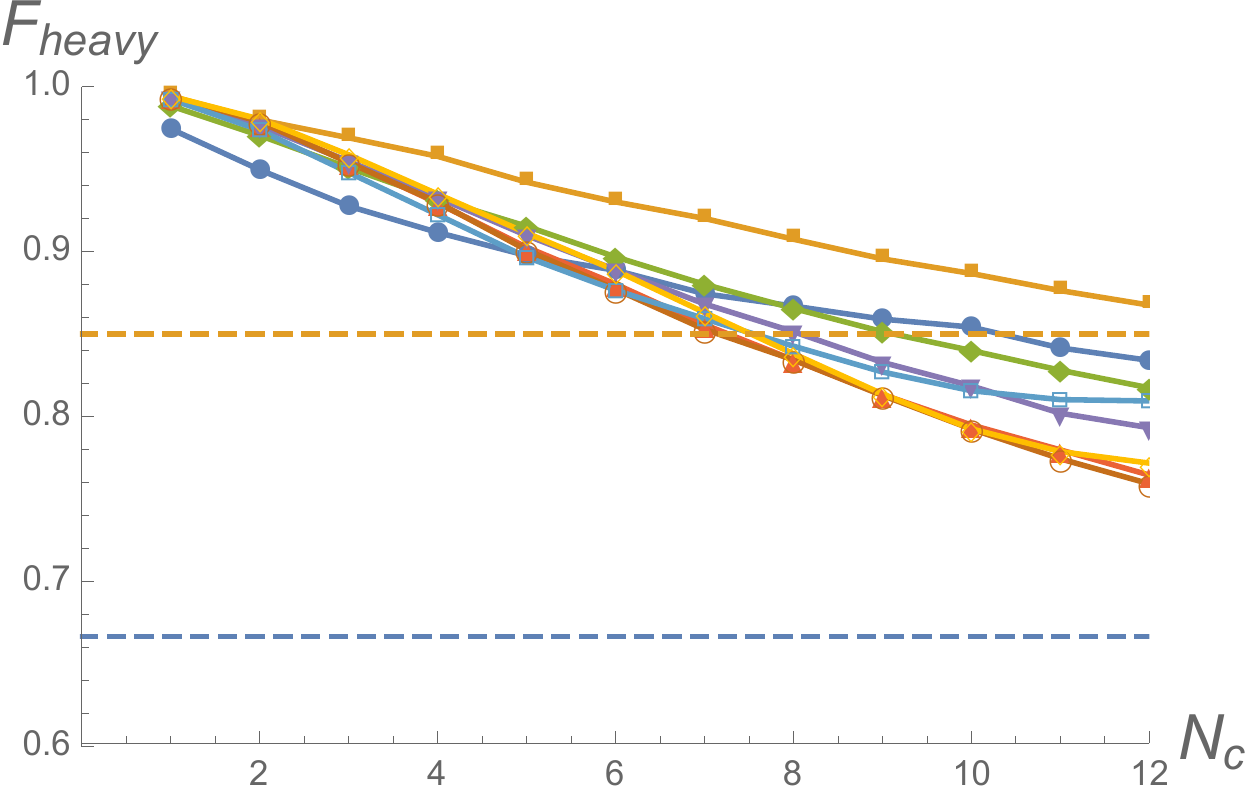}

\includegraphics[width=2.25in]{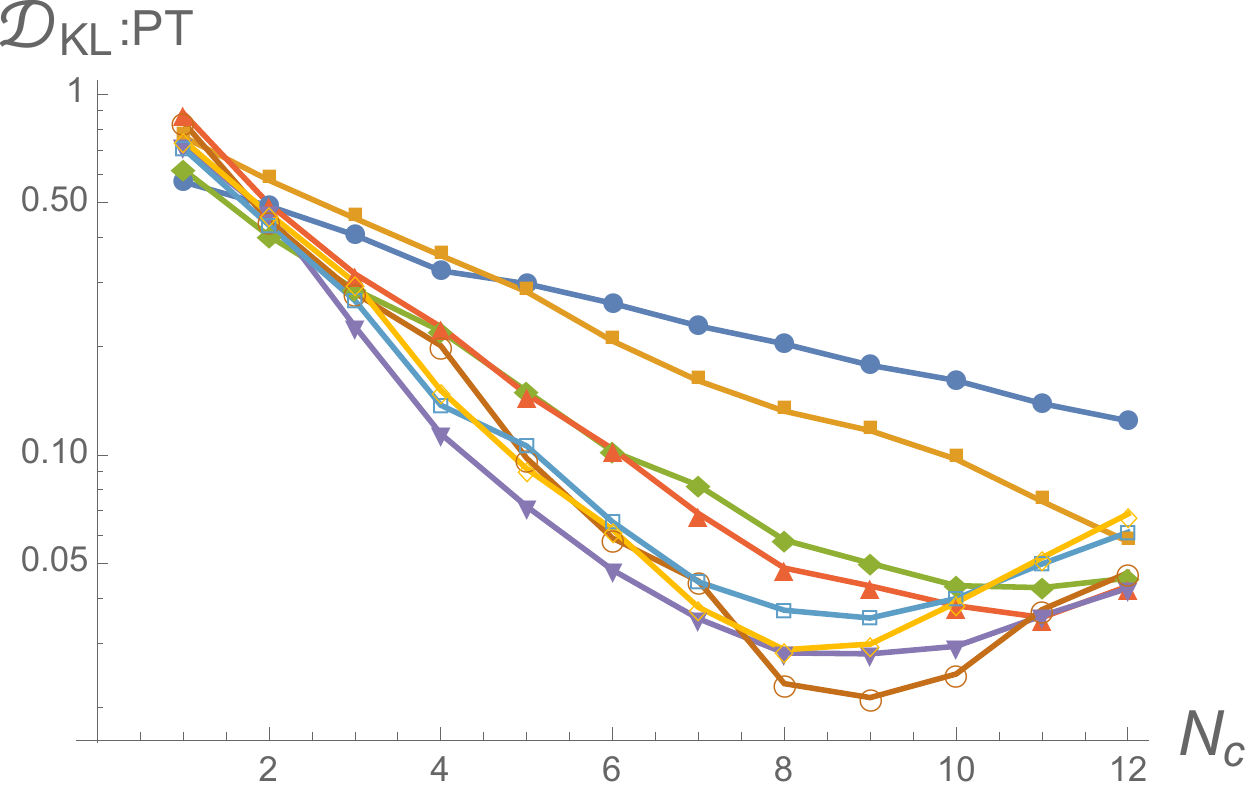}
\includegraphics[width=2.25in]{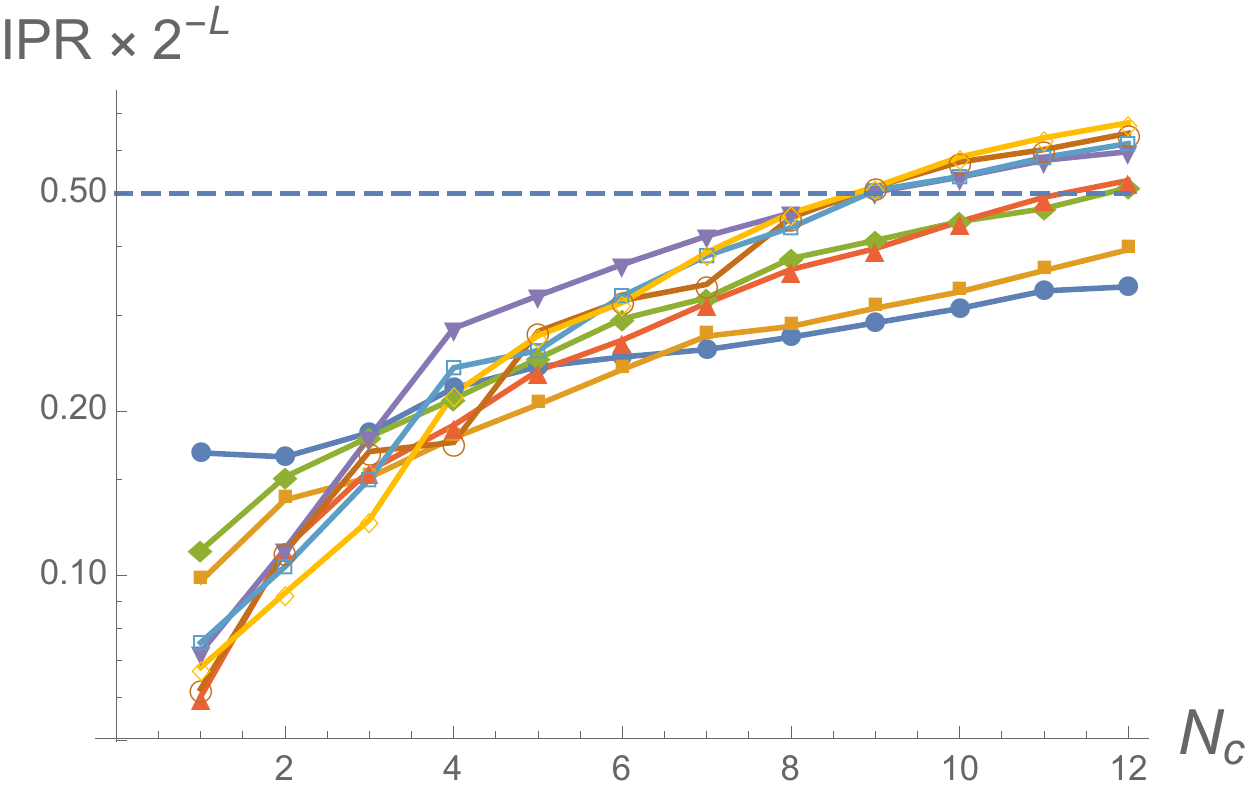}
\includegraphics[width=2.25in]{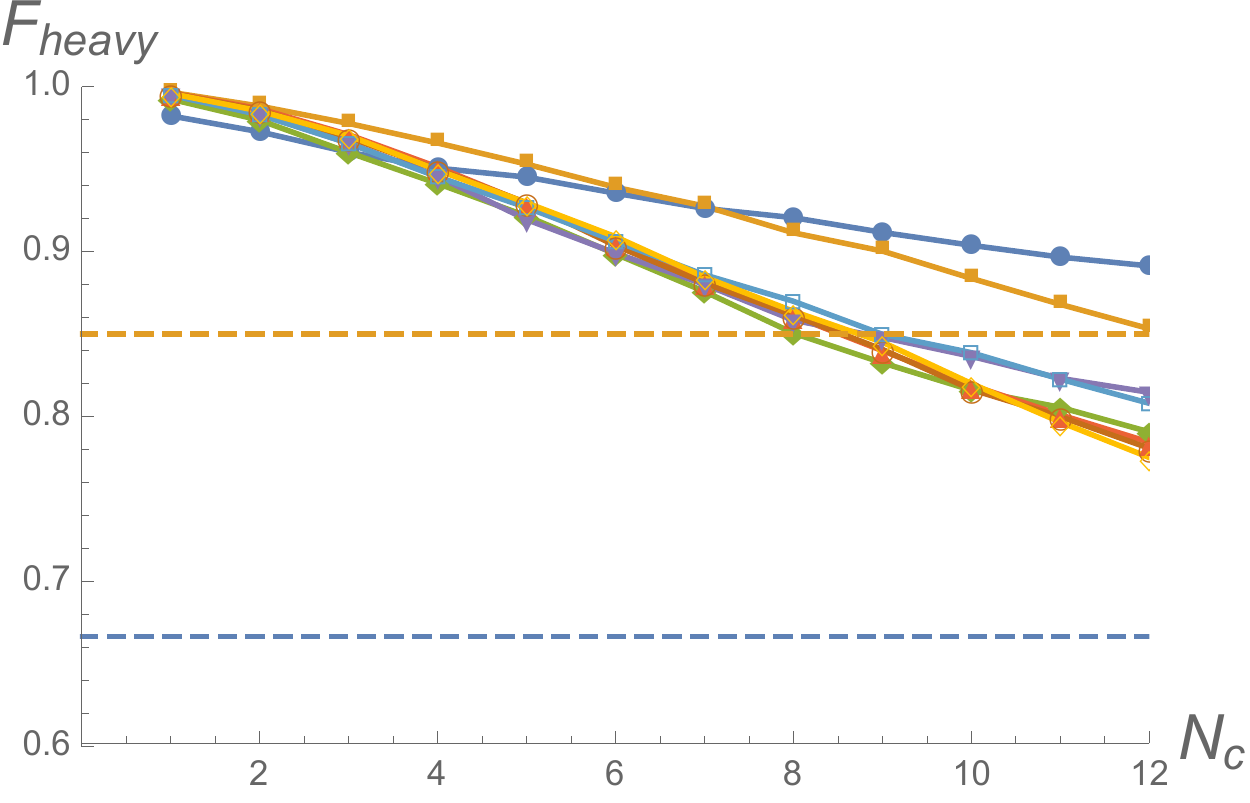}
\caption{Output statistics, showing significant information scrambling. In this and all subsequent figures unless otherwise noted, $L=4$ is plotted with blue filled circles, $L=5$ gold boxes, $L=6$ green diamonds, $L=7$ red triangles, $L=8$ purple triangles, $L=9$ brown open circles, $L=10$ light blue open squares and $L=11$ yellow open diamonds. The top row corresponds to simulations in parametrization A, the bottom row the same quantities in parametrization B. Left: Kullback-Leibler divergence from an ideal Porter-Thomas distribution over the entire qubit Hilbert space, as a function of the number of cycles $N_c$ of evolution. A Porter-Thomas distribution is a key signature of quantum chaotic evolution; in our protocol both parametrizations come very close to such a distribution, with an average minimum K-L divergence of around 0.02, before slowly pulling away from one at long times as a likely trivial final state is approached (the time to reach such a state is expected to be many times longer than the window shown here). Center: inverse participation ratio (IPR) vs $N_c$. Consistent with the Porter-Thomas output and volume-law entanglement, the IPR measurement shows that the system explores a constant fraction of its total Hilbert space as $L$ grows, demonstrating that an exponentially large amount of classical information is required to represent the state after just a few cycles of evolution. Right: fraction of sampled bit strings which are ``heavy," e.g. larger than the median output probability. Aaronson and Chen \cite{aaronsonchen2016complexity} have argued that a sufficient fraction, for example 2/3 (blue dashed line) is a strong indicator of classical intractability for random quantum circuits; a Porter-Thomas distribution produces heavy output in approximately 85\% of samples (gold line). All of our simulations are well above the 2/3 threshold even at fairly long evolution times.}\label{outputfig1}
\end{figure*}


\begin{figure}
\includegraphics[width=3.25in]{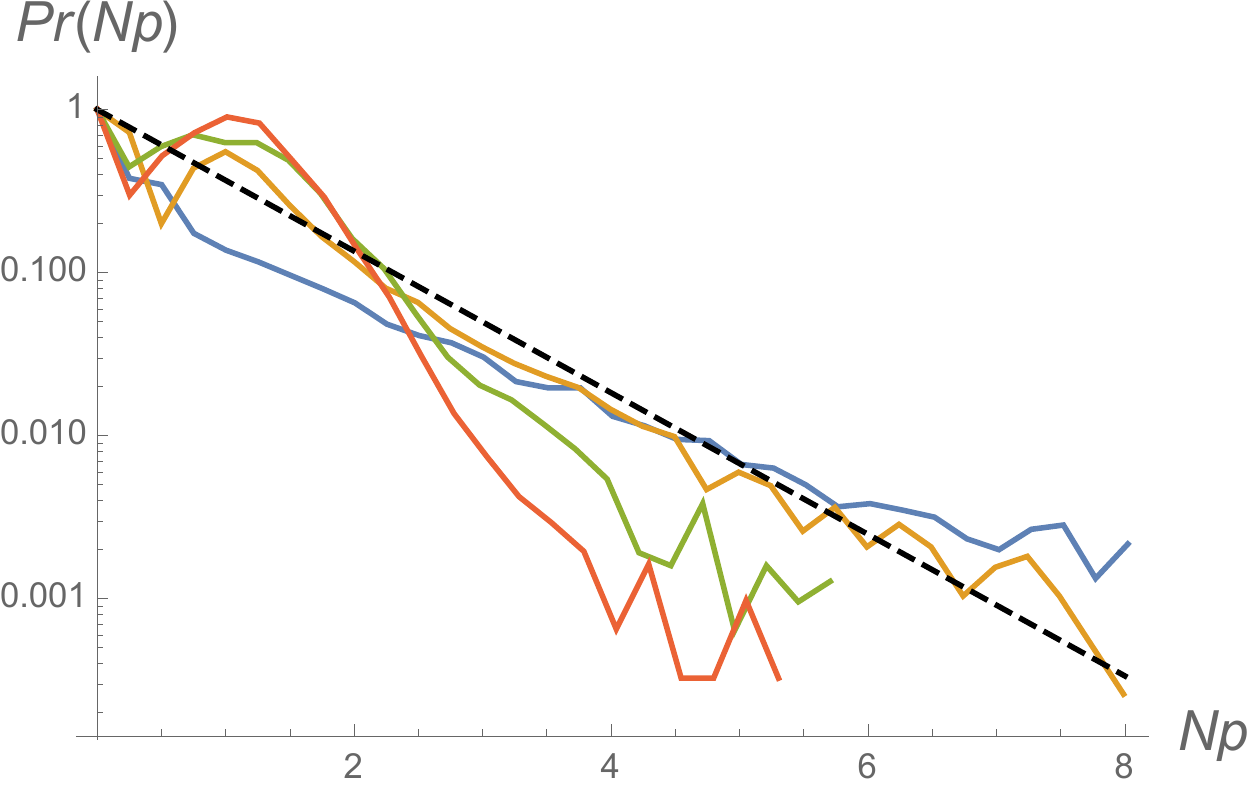}
\caption{Distribution of output probabilities for parametrization A, with $L=9$, after three (blue), six (gold), nine (green) and twelve (red) cycles of evolution, combining the results of 32 protocol instances for a total of 16384 data points per curve. Here, $N = 2^L$ is the qubit Hilbert space size, and the plotted quantity is the average probability of a given configuration having probability $p$ in the final output distribution (note that the $x$ axis is rescaled by a factor of $N$). The black dashed line, $e^{- N p}$, corresponds to an ideal Porter-Thomas distribution, the result of fully chaotic quantum evolution. At six cycles the distribution is very close to P-T, with an average K-L divergence of 0.02 from an ideal P-T distribution, but for longer evolutions the system pulls away from it toward a distribution closer to, but clearly distinct from, incoherent uniform randomness. See the main text and FIGs.~(\ref{outputfig1},\ref{outputfig2}) for more details.}\label{odistfig}
\end{figure} 

\begin{figure*}
\includegraphics[width=3.25in]{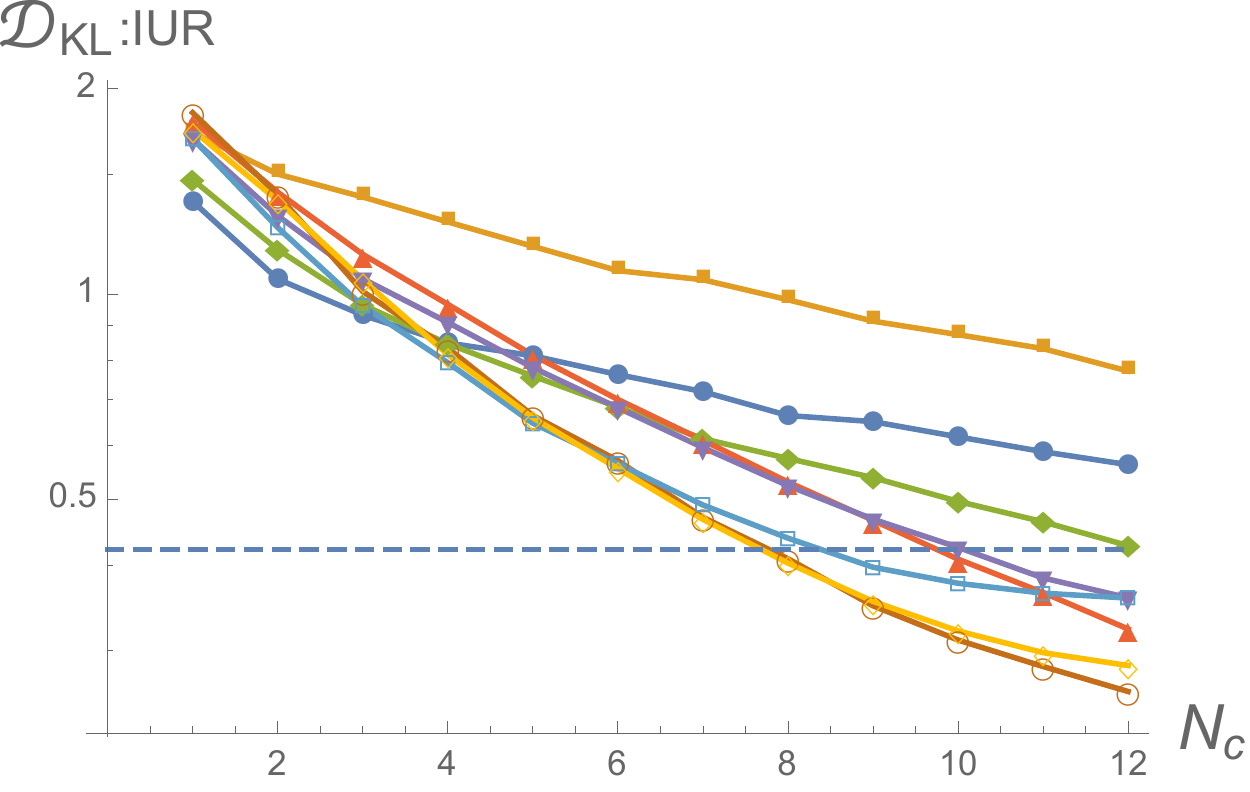}
\includegraphics[width=3.25in]{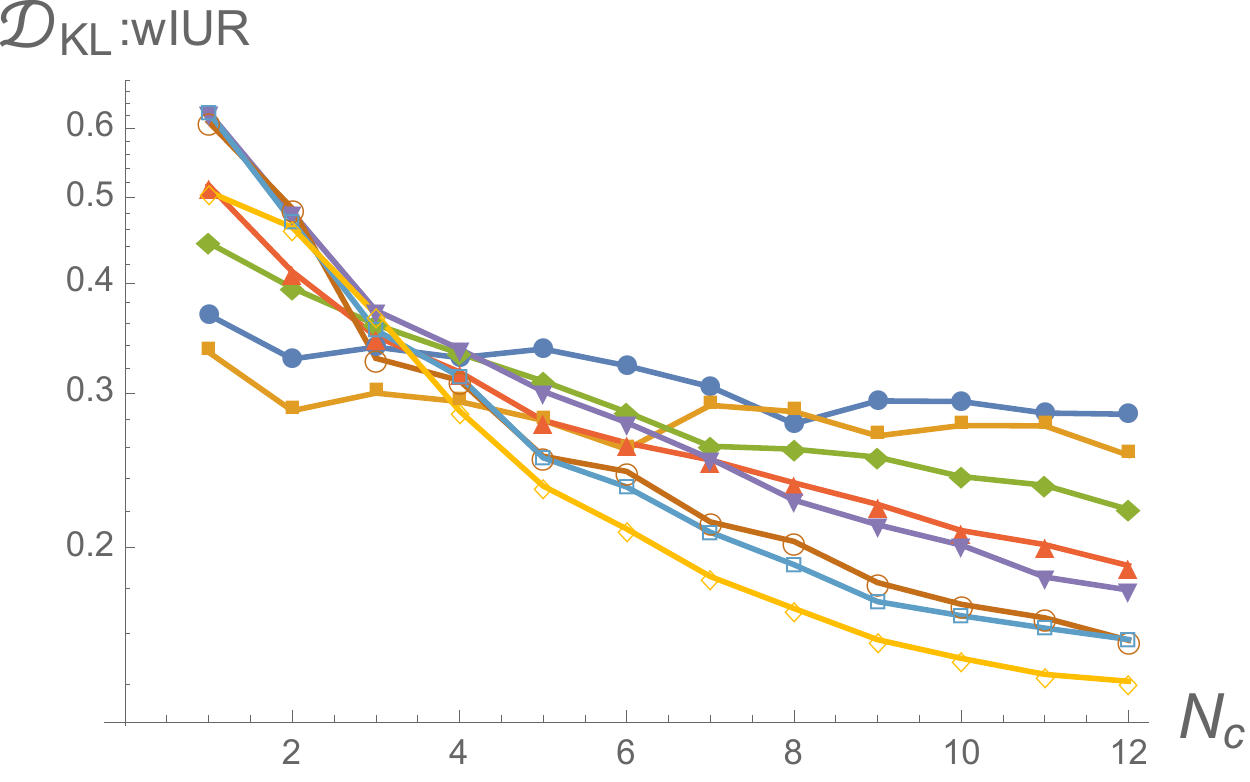}

\includegraphics[width=3.25in]{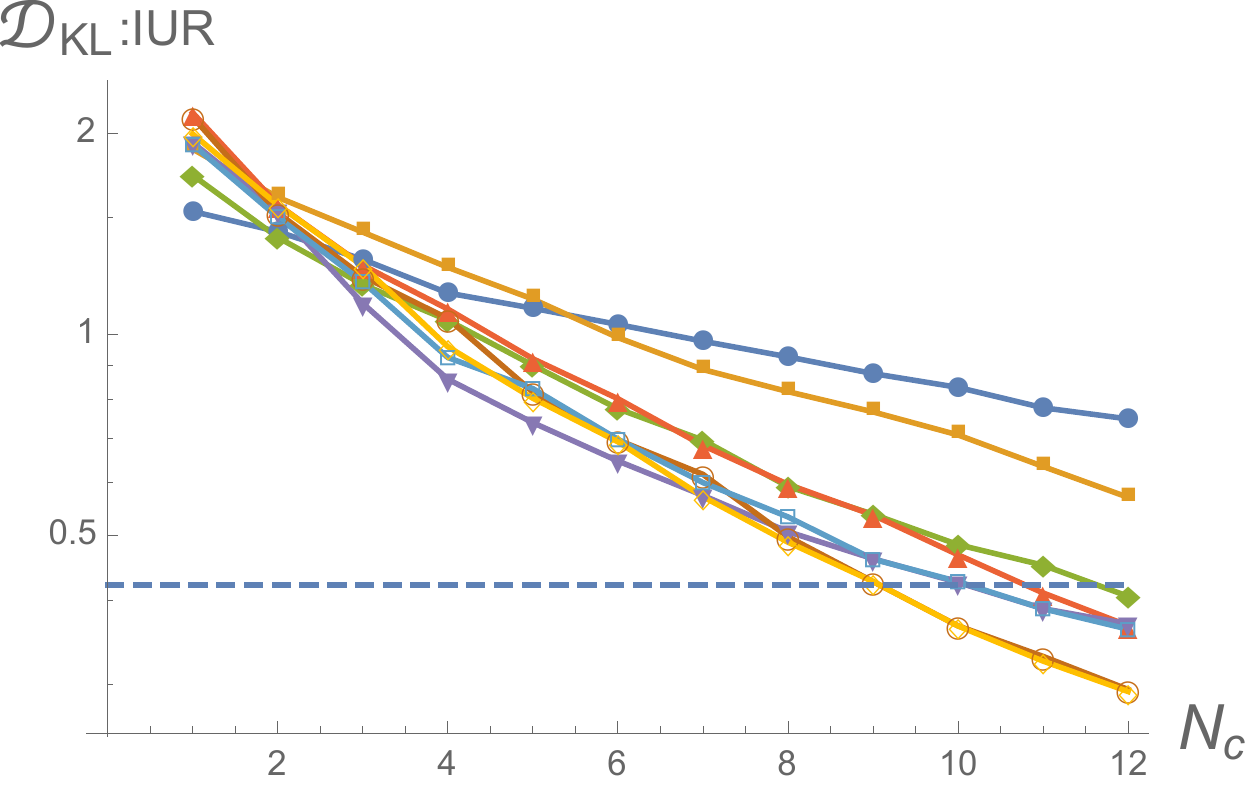}
\includegraphics[width=3.25in]{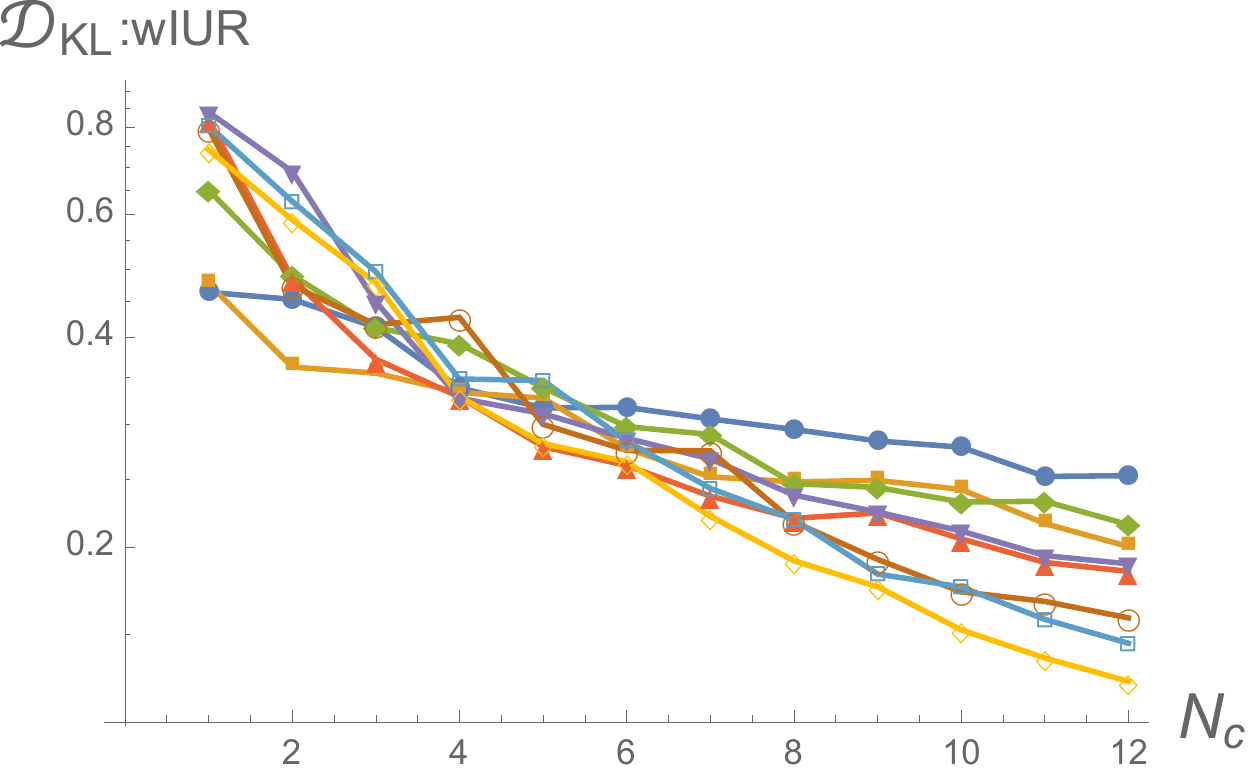}
\caption{Output statistics, showing that simple classical distributions do not capture the output of our system. The top row corresponds to simulations in parametrization A, the bottom row the same quantities in parametrization B. Left: K-L divergence from incoherent uniform randomness (all $P_k = 2^{-L}$), the result of noise in a random quantum circuit; such a distribution does not well approximate our system even when photon loss has become significant. The dashed line at $\sim 0.422$ is the K-L divergence between IUR and an ideal Porter-Thomas distribution. Right: K-L divergence from a reweighted variant of IUR, where relative probabilities are Poisson-weighted (see Eq.~\ref{defWIUR}) by the total number of added particles beyond the population of the initial state; while this distribution is a better representation than pure IUR, it still does not capture the complex quantum structure generated in evolution.}\label{outputfig2}
\end{figure*}

\begin{figure*}
\includegraphics[width=2.25in]{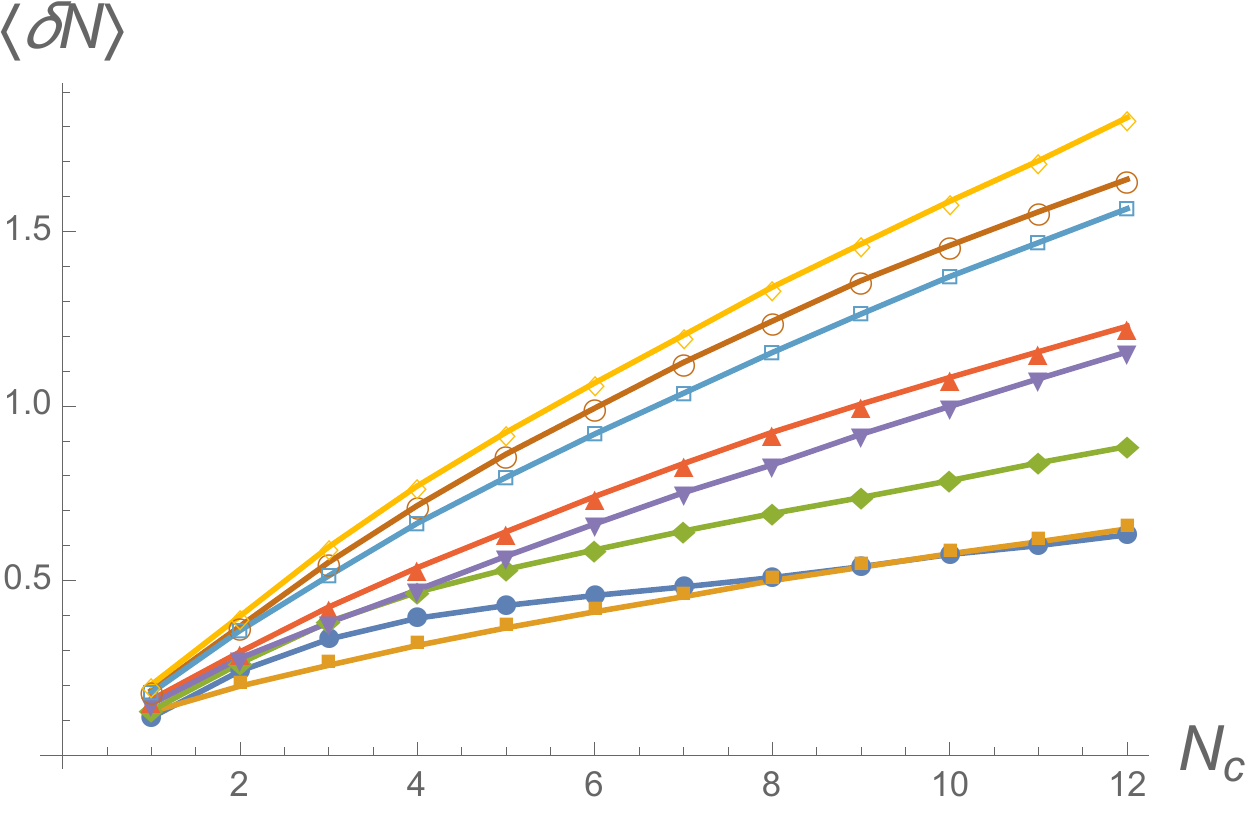}
\includegraphics[width=2.25in]{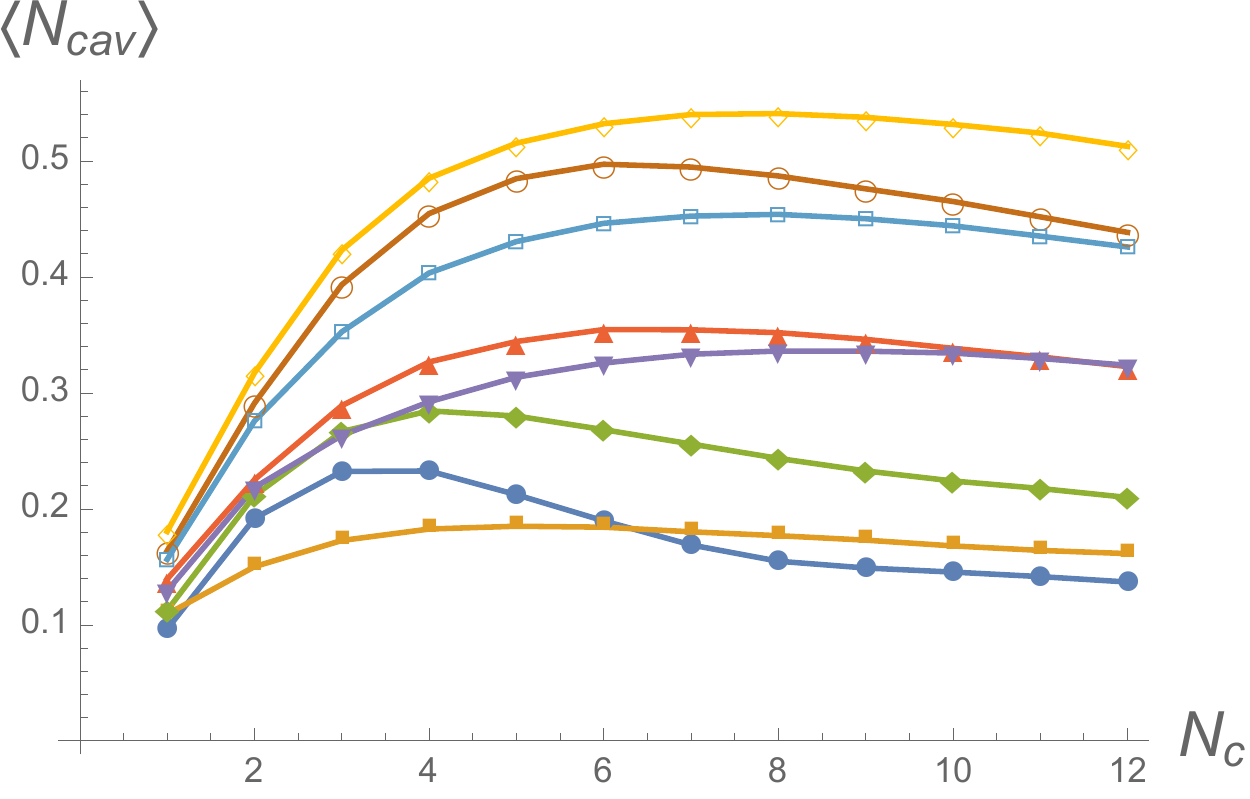}
\includegraphics[width=2.25in]{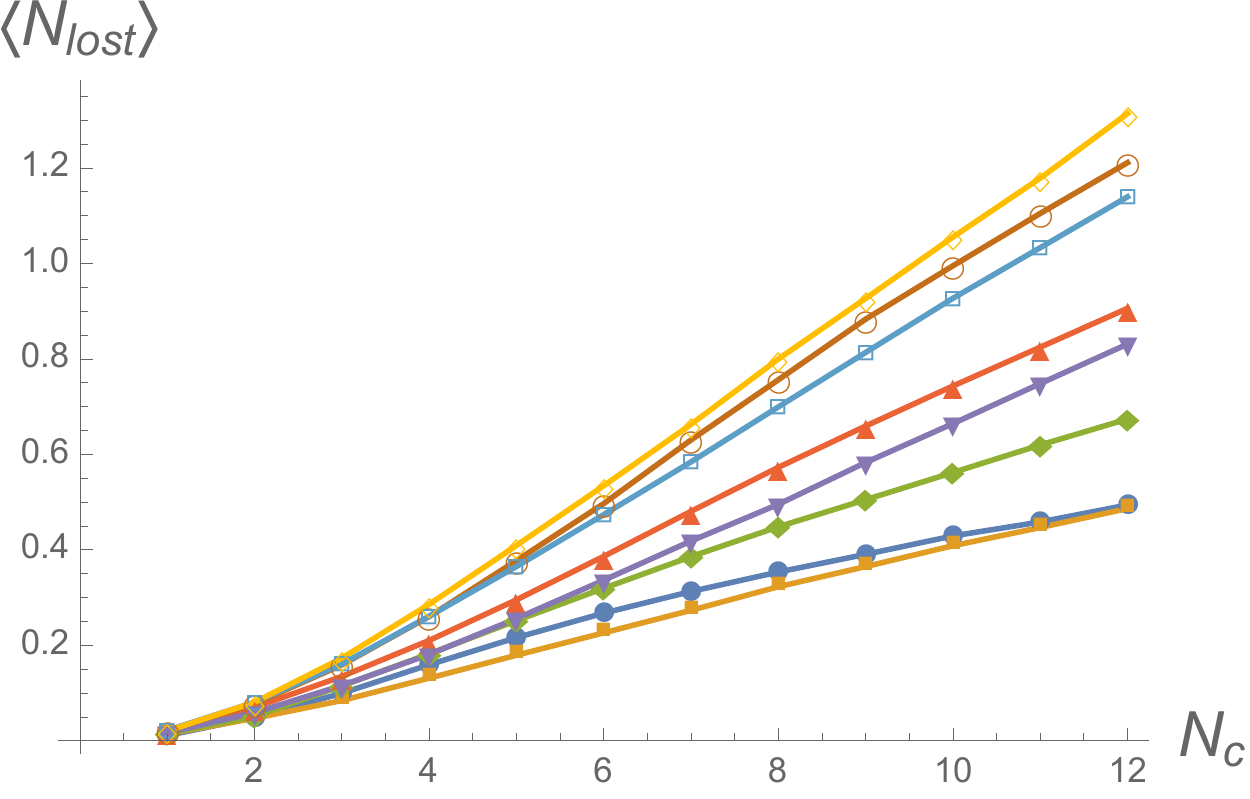}

\includegraphics[width=2.25in]{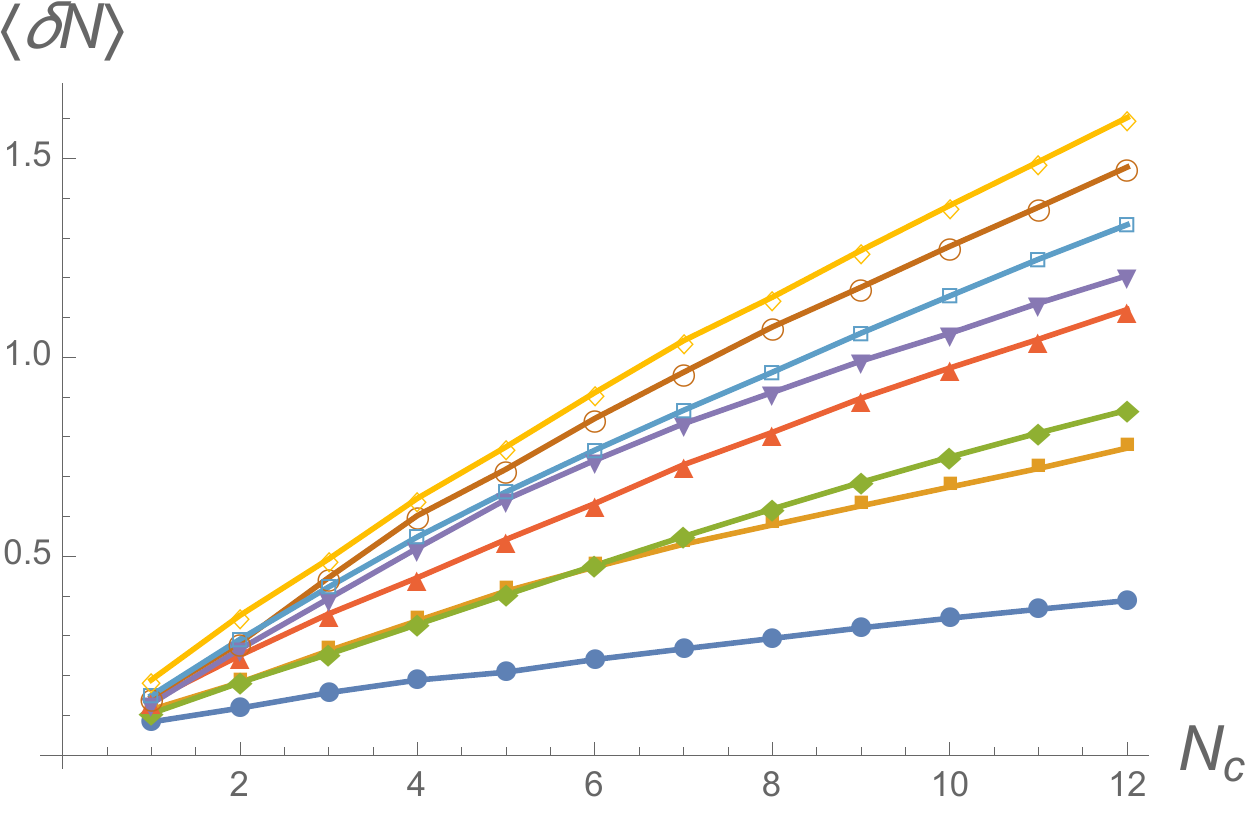}
\includegraphics[width=2.25in]{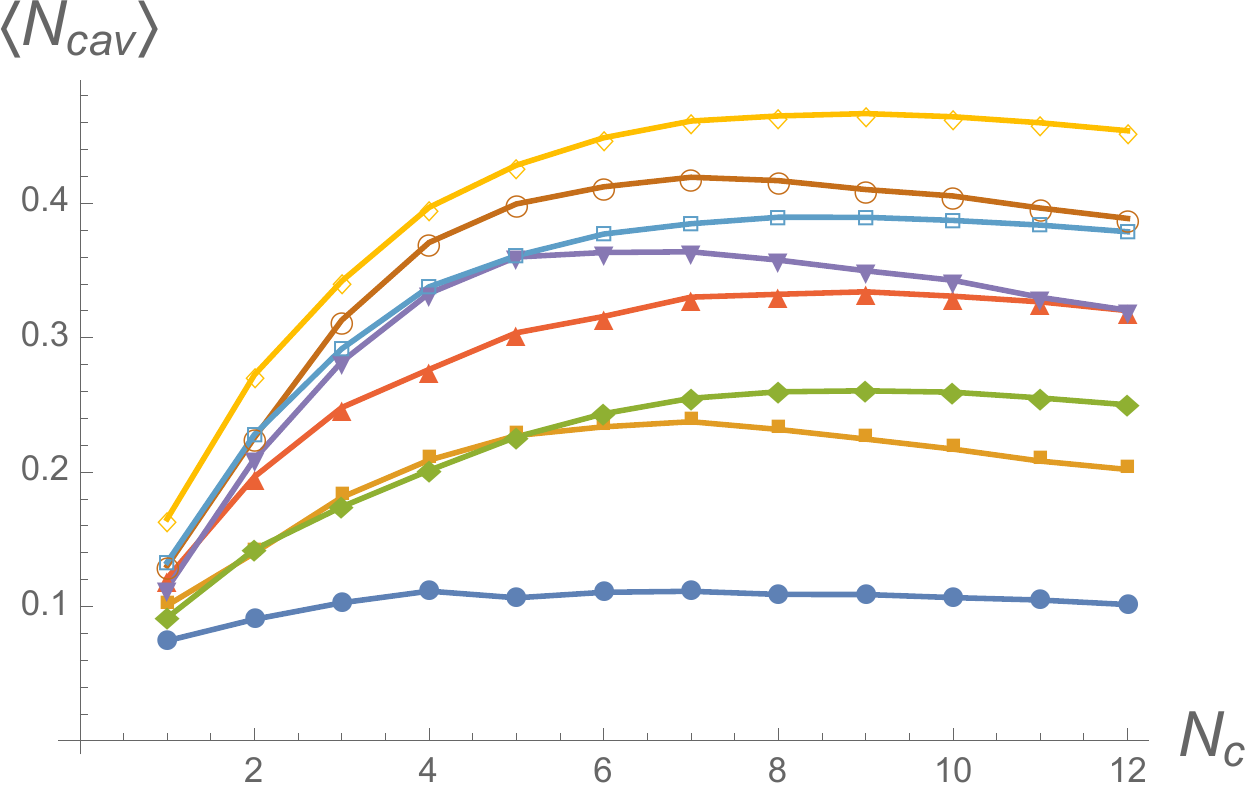}
\includegraphics[width=2.25in]{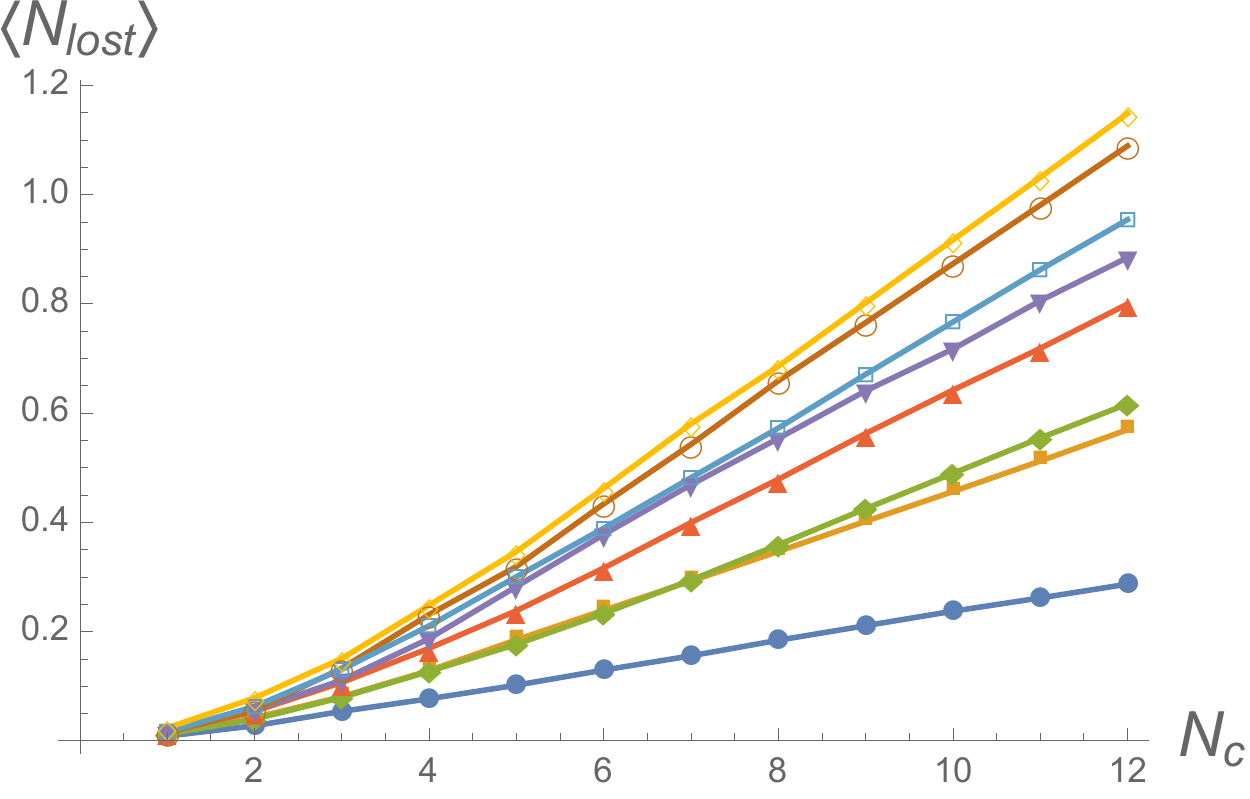}
\caption{Number fluctuations, cavity photon populations, and cavity photon loss. The top row corresponds to simulations in parametrization A, the bottom row the same quantities in parametrization B. Left: average number of photons added to the qubits vs $N_c$, which grows extensively with $L$, though with significant even-odd effects in parametrization A (see text for details). Center: average total cavity photon population, which also grows extensively, though with a small prefactor due to a combination of relatively weak coupling to resonant modes in the qubit chain and the significant loss rate. This quantity has important implications for the simulation difficulty; the larger it is, the more cavity photon states need to be included in the classical Hilbert space for faithful simulations. Right: average number of cavity photons lost. This quantity is $O \of{1}$ near the end of the evolution time, showing that the effects of noise cannot be ignored, but do not trivialize the dynamics, in our evolving chain.}\label{nqnlfig}
\end{figure*}

Having thoroughly studied entanglement generation and loss in our noisy system, we now examine the output distribution itself. To do so, we use the familiar Kullback-Leibler divergence \cite{kullbackleibler1951} to quantify the ``distance" between our observed output distribution and other important ones:
\begin{eqnarray}\label{defKL}
\mathcal{D}_{KL} \of{\rho_{A},\rho_{B}} &\equiv & \sum_{i} P_{Ai} \ln \frac{P_{Ai}}{P_{Bi}}.
\end{eqnarray}
In FIG.~\ref{outputfig1}, we plot the K-L divergence of the full output distribution in the qubit basis from a Porter-Thomas (P-T) distribution, as a function of the number of cycles of evolution, averaged over random instances of each protocol. The P-T distribution used for comparison is defined over the full $2^L$-element qubit Hilbert space, and not a restricted subspace as in the unitary protocol which conserves photon number. Consistent with quantum chaotic behavior at intermediate times, the output distribution becomes very close to a P-T distribution between 6 and 9 cycles of evolution (for the simulation parameters chosen, and as seen in the figure, this is somewhat protocol dependent) before gradually pulling away at longer times; see FIG.~\ref{odistfig} for example output distributions. Note that since the point of ``closest approach" varies from instance to instance the averages plotted here tend overestimate the minimum distance achieved for a given instance. 

What is rather remarkable about these results is that cavity photon losses are already significant (see FIG.~\ref{nqnlfig}) by the time a P-T distribution well fits the observed output, with (for $L=9$) an average of $\sim 0.9$ photons lost by 9 cycles for parametrization A and $\sim 0.75$ photons lost by 9 cycles for parametrization B. As discussed below, this signature of quantum chaos is \textit{not} observed when considering random incoherent processes in the \textit{qubits}, which rapidly drives the system toward trivial configurations and cannot generate new correlations. Viewed alongside the persistence of entanglement after a photon loss discussed in the previous section, these results confirm that photon loss from a resonantly coupled auxiliary system is qualitatively different from random qubit error, and leads to highly nontrivial quantum dynamics.

However, as shown in FIG.~\ref{outputfig2} there is some ``trivializing" effect to the cavity photon loss, in that the observed distribution grows closer to incoherent uniform randomness (IUR) at long times (before eventually reaching a fully occupied lattice at extremely long times, assuming that no photon loss processes balance out the blue sideband terms), consistent with a trivial final state. Given effort to tailor the protocol to stabilize nontrivial configurations at long times (see for example \cite{kapithafezi2014,masaxberg2018dissipatively}), we would expect this effect to disappear, but such considerations are beyond the scope of this work.

Importantly, in both sets of trials (though much more pronounced in parametrization A), there are clear even-odd effects; odd $L$ cases have higher values for peak entanglement, number fluctuations, and average cavity photon population (and thus, loss rates). The reason for this likely comes from the choice of cavity detuning-- in parametrization A, the cavity detuning $h_{Ci}$ in Eq.~(\ref{HQC}) is set to zero, whereas all the $h_{Ci}$ are assigned random single photon hopping energies in parametrization B. As remarked earlier, since $\Omega_{QC,max} \ll g_{max}$, a photon can only be added or removed from the chain if it populates a near-resonant propagating mode, and when we consider the eigenvalues of a single particle hopping on a 1d chain with open boundary conditions, there is a zero energy mode for odd $L$, but not for even $L$. Thus, while this simplistic picture is complicated by interactions, disorder, and the qubit-cavity dispersive shift, it is reasonable to assume that the odd $L$ chains are \textit{on average} closer to resonance with the cavities than the even $L$ chains, and thus interact with them more strongly. Further, since the density of states of the interacting system peaks at the center of the spectrum, we expect some enhancement for odd $L$ even in parametrization $B$, where a single particle tunneling energy lines up with the peak. This explains why odd $L$ chains have larger peak entanglement, fluctuations and cavity loss rates than even $L$ chains do, though we expect this effect to diminish as $L$ becomes large.

Further, as shown in FIG.~\ref{outputfig1}, we also computed the inverse participation ratio (IPR), and as is to be expected from our previous results, our protocol explores an $O\of{1}$ fraction of Hilbert space, typically reaching half of the maximum value of $2^L$ between 6 and 10 cycles, depending on protocol details. Combined with the exponentially growing entanglement negativity and the lack of any symmetries to exploit, an exponential amount of classical information is thus required to exactly store the evolving quantum state.

\subsection{Output heaviness}

Recently, Aaronson and Chen provided an alternative metric for quantum sampling hardness, called heavy output generation (HOG) \cite{aaronsonchen2016complexity}. The HOG problem is stated as follows: given a suitably randomized quantum circuit, generate an output distribution for which at least two thirds of the observed samples $\cuof{x_1, ... , x_N}$ have a higher probability than the median value of all probabilities $\cuof{P_k}$ in the full output distribution. Aaronson and Chen proved that if a plausible conjecture called QAUTH is true, no polynomial-time classical algorithm can solve the HOG problem in the most general cases. Note that for a Porter-Thomas distribution, approximately 85\% of the sampled outcomes will have greater than median probability, so a perfectly executed random quantum circuit or unitary Bose-Hubbard evolution easily satisfies the heavy output criteria. Conversely, an RQC executed with poor fidelity produces a distribution very close to IUR, and does not satisfy the heavy output criteria, though it may still be exponentially difficult to reproduce classically. 

In practice, a heavy output distribution is not completely sufficient to prove classical hardness, given that classically easy examples, such as low-depth circuits or ones composed entirely of Clifford gates, can also have heavy output distributions. However, absent any obvious simplifying factors, heavy output can be a valuable metric for classical difficulty \cite{crossbishop2018}, so it is reasonable to check if our simulations produce it\footnote{Formally, the hardness proof for HOG assumes the output distribution is generated by a random quantum circuit, and while instances of our protocol can of course be represented as a subset of that family given that time evolution can be Trotterized and non-unitary operations can be modeled through coupling to additional ancillary qubits, the constraints on randomness that result would make it very much an edge case. It is thus possible that the HOG hardness proof could be shown to not apply to our system, though we nonetheless consider heavy output in our protocol to further bolster our arguments for classical simulation difficulty.}. In FIG.~\ref{outputfig1} we plot the heavy output fractions observed in both parmetrizations; all simulations show an output heaviness substantially greater than 2/3. These results clearly demonstrate that our protocols satisfy the heavy output criteria, bolstering our expectations for classical difficulty. When combined with volume-law entanglement scaling, full Hilbert space exploration, output distributions showing signatures of quantum chaos, high effective circuit depths (see the section on classical difficulty for more details), and the lack of any symmetries to simplify the evolution, we find it extremely doubtful that any polynomial-time classical algorithm could reproduce our results once $L$ becomes large.

\subsection{Fidelity loss from qubit error}

To discuss the effect of noise we must first define a fidelity metric. Throughout this work we will use a simple, and experimentally relevant, definition of fidelity based on the K-L divergence described above:
\begin{eqnarray}\label{defKLfid}
\mathcal{F} \of{P_{obs}} & \equiv & 1 - \frac{\mathcal{D}_{KL} \of{P_{ideal},P_{obs}} }{\mathcal{D}_{KL} \of{P_{ideal},P_{TC}}}.
\end{eqnarray}
Here, $P_{ideal}$ is the probability distribution of a perfectly executed instance of the protocol, $P_{obs}$ is the observed result of the experiment (likely including noise), and $P_{TC}$ is a trivial classical distribution, the choice of which depends on protocol details. While this does not coincide with the standard definition of fidelity, it captures a notion of  statistical distance. Note that while for RQC the choice of trivial distribution is not fundamental~\cite{boixoisakov2016}, the most convenient is incoherent randomness (IUR), where all $P_i = 1/N_A$ for an output space of dimension $N_A$. For the unitary protocol initialized with $N_{ph}$ photons in the qubits, $N_A = {L \choose N_{ph}}$. In cases where $\mathcal{F}$ falls below zero, we assume it to be zero; for a Porter-Thomas distribution, $\mathcal{D}_{KL} \of{P_{PT},P_{IUR}} = 1 - \gamma \simeq 0.423$, where $\gamma$ is the Euler-Mascheroni constant. The choice to normalize the K-L divergence based on the divergence from trivial classical distributions is motivated by the empirically observed results from RQC, where IUR is the distribution that results from one or more Pauli errors occurring during the evolution, thus sending $\mathcal{F}$ to zero. It also in some sense measures performance above a trivial classical result; since simulating the system's evolution with an IUR distribution is computationally ``free" it makes sense to let that level of accuracy be zero fidelity, and let nonzero fidelities thus correspond to better approximations of the intended quantum dynamics. Note that when studying fidelities for the intentionally noisy protocol that we focus on in this work, the ideal simulation $P_{ideal}$ includes the \textit{intentional} noise sources $\cuof{O_i}$ (in our case, cavity photon loss), but not unintentional ones (control errors in the operations, phase and loss errors in the qubits, and so forth). 

\begin{figure}
\includegraphics[width=3.25in]{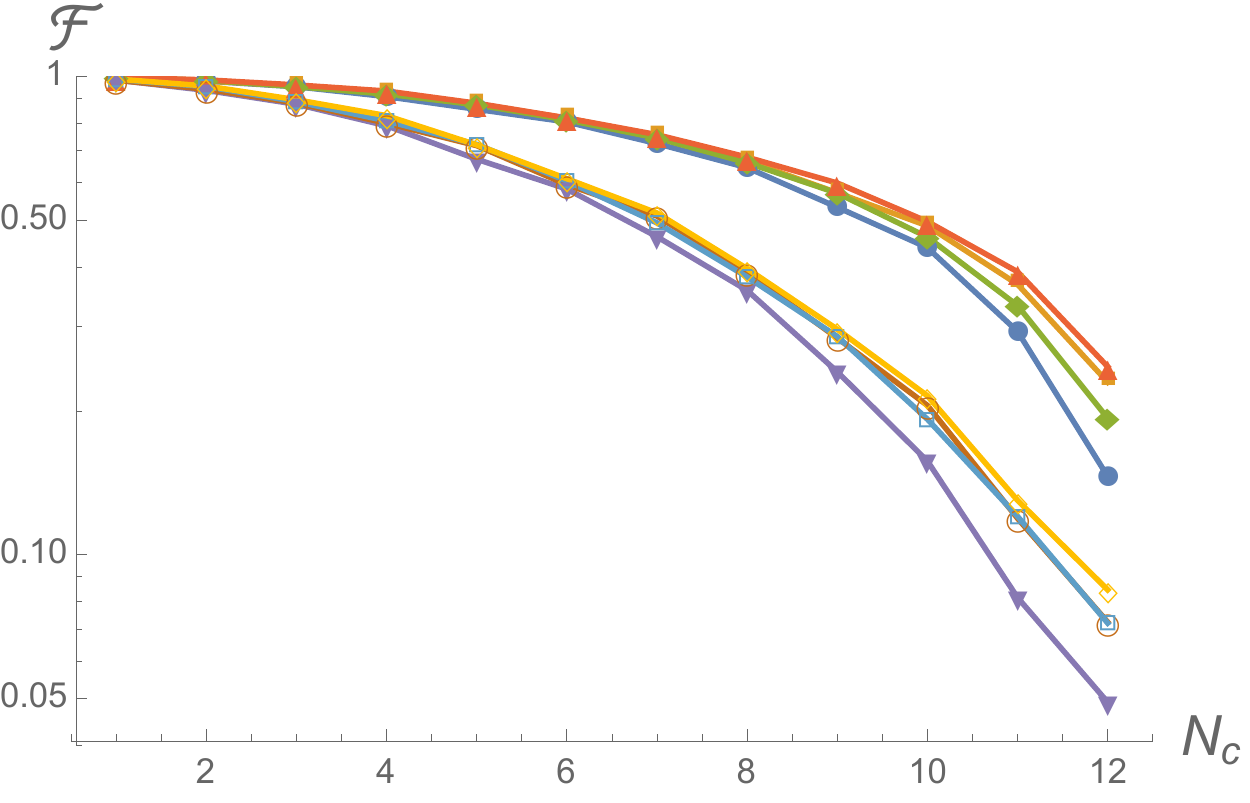}
\includegraphics[width=3.25in]{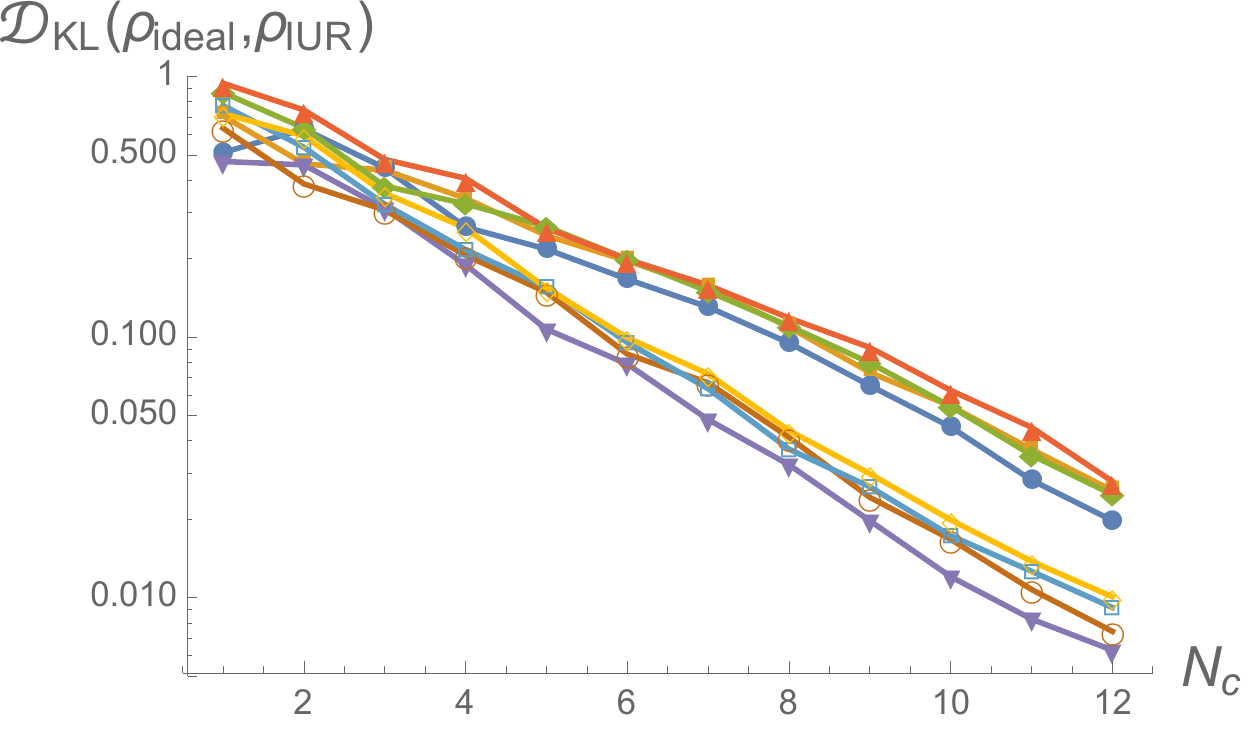}
\caption{Fidelity (top) and K-L divergence from incoherent uniform randomness (bottom) for phase errors in the purely unitary protocol with no qubit-cavity interactions, for $L$ running from 6 to 9. The colors used in this plot differ from other figures in this work-- in the top cluster of curves, $L=6$ is blue, 7 is gold, 8 is green and 9 is red. In the bottom cluster, 6 is purple, 7 is brown, 8 is light blue and 9 is yellow. In the higher (in fidelity and divergence from IUR) clusters of curves we average over a single phase error insertion during 12 cycles of evolution, and the lower clusters of curves correspond to averaging over two random phase error insertions. Somewhat surprisingly, the fidelity loss from a phase error is highest for $L=6$ and decreases slightly as $L$ increases toward 9, though this effect would be swamped by the linearly increasing rate of errors with $L$ in a real experiment. As shown in the second plot, the output distribution averaged over error insertions is difficult to distinguish from incoherent uniform randomness, where all states with the appropriate total photon number have equal probability. A single photon loss error sends $\mathcal{F}$ to zero.}\label{fidUfig}
\end{figure}

To ground our results, we first consider random qubit error, in the form of white noise phase errors and $T_{1}$ photon loss, applied to the unitarily evolving chain with no qubit-cavity interactions. Since the applied Hamiltonian conserves total photon number, a single photon loss instantly sends the fidelity to zero, though we can eliminate these events, as well as most SPAM errors, through post-selection since any change in total photon number implies an error has occurred. Random photon addition has the same effect, though this is an empirically much weaker noise channel in superconducting qubits. Phase noise, on the other hand, is not detectable, and reduces the fidelity significantly, though unlike RQC a single error does not appear to send $\mathcal{F}$ strictly to zero; as shown in FIG.~\ref{fidUfig} averaging over the insertion of a single error leads to $\mathcal{F} \simeq 0.25$ after 12 cycles of evolution for the parameters described above. Averaging over two error insertions gives $\mathcal{F} \simeq 0.076$, a further reduction by a factor of 3.3, suggesting that fidelity decreases exponentially with the number of phase errors, as we would expect for a system with chaotic dynamics. If we use an alternative fidelity measure based on the absolute distance $\mathcal{D}_{Abs}  \of{P_A, P_B} \equiv \frac{1}{2} \sum_i \abs{P_{Ai} - P_{Bi} }$ we find somewhat smaller final fidelities for one and two phase errors, but in both cases $\mathcal{F}$ remains nonzero\footnote{We hypothesize that a small but nonzero fidelity persists due to the structure of the many-body ``gate," where all qubit-qubit exchange couplers are turned on simultaneously. Since the qubit nonlinearity strongly, though not completely, suppresses double occupancies, spin configurations where many sites in a row are all occupied by photons are less likely to be produced by applications of the coupler pulse than ones where occupied and empty sites alternate. Consequently, even in cases where phase errors have occurred, the same relative biases away from particular classes of states apply, and the divergence between the error trajectories and the ideal ones is slightly lower than between the ideal trajectory and IUR. Note that if this hypothesis is correct, we expect its effect to be diminished in 2d, where photons cannot blockade each others' motion to the same extent, and residual fidelities after phase errors will likely be closer to zero.}. Note that since phase errors are along the directions of both the initial product state and final qubit measurement, they can only influence the output distribution through changing the result of subsequent coupler pulses, and thus have less influence at short times. This effect can be seen in real experimental data (see figure 4 of \cite{neillroushan2017}), and in numerical simulations; we found that averaging a single phase error over just three cycles instead of twelve leaves a final fidelity of approximately 0.5, twice the fidelity obtained when averaging over a single error occurred in twelve cycles of evolution.

We find similar results in our noisily evolving chain, though care must be taken in defining a fidelity metric in that case, due to the non-conservation of photon number. In our noisily evolving chain with incoherent (if quantum correlated) particle addition, we can better approximate the final distribution with a re-weighted modification of the IUR distribution, which we call WIUR, where the individual bit string probabilities are reweighted by a function of their total qubit photon number, assuming a Poisson distribution of random addition or loss events starting from the known initial photon number. WIUR is also computationally trivial distribution, and like IUR it does not accurately capture the output distribution of our protocol, but \textit{does} provide a somewhat better approximation to the full system dynamics than IUR over the full $2^L$-element qubit Hilbert space. For concreteness, assume the system begins with $N_0$ photons in the qubits, and an average of $\delta N$ photons are added after $N_c$ cycles of evolution\footnote{While qubit $z$ errors will scramble the relative amplitudes of states within a given band of fixed photon number, we do not expect them to significantly change the \textit{average} number of photon creation or loss events induced by the cavities. This statement assumes that the positions and times of $z$ errors are being averaged over, and may not be the case for comparing the full output distribution of an error-free protocol instance with one where one or more $z$ errors occur at specific point(s) in spacetime.}. We then generate the WIUR distribution by assigning all bitstrings $\ket{k}$ probabilities given by:
\begin{eqnarray}\label{defWIUR}
P_k &=& 0 \; \; \; \cuof{N_k < N_0}, \\
&=& \frac{1}{W} {L \choose N_k }^{-1} e^{-\delta N}  \frac{  \of{\delta N}^{\of{N_{k}- N_{0}}}   }{\of{N_k- N_0}!}  \; \; \; \cuof{N_{k} \geq N_0}. \nonumber
\end{eqnarray} 
Here, $N_k$ is the number of photons in bitstring $\ket{k}$ and $W$ is a normalization factor such that $\sum_k P_k = 1$. Analogues of this distribution can be easily defined for other protocol choices. Using this distribution to replace the IUR distribution in (\ref{defKLfid}), we can then estimate reductions to $\mathcal{F}$ by averaging the evolution of a given protocol over the insertion of a single photon loss error, or one or two phase errors as in the unitary protocol above.

\begin{figure}
\includegraphics[width=3.25in]{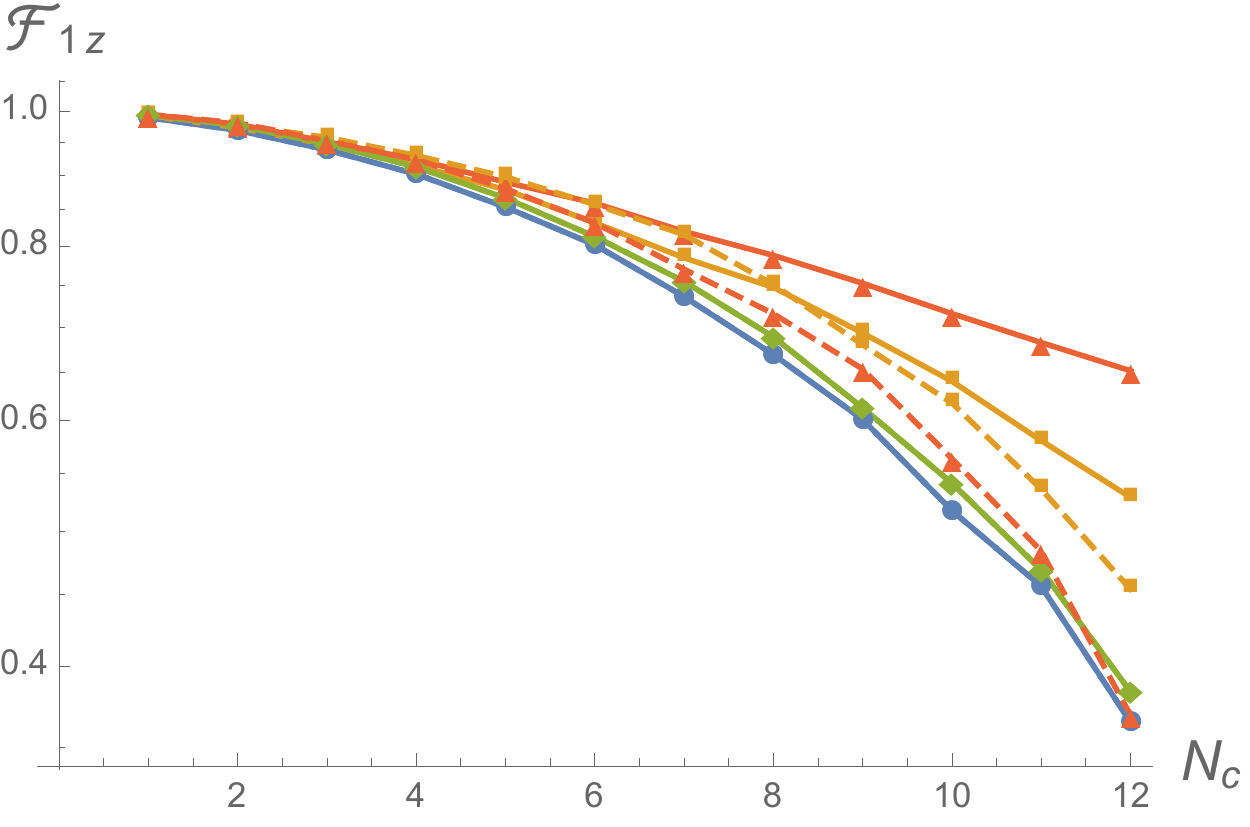}
\includegraphics[width=3.25in]{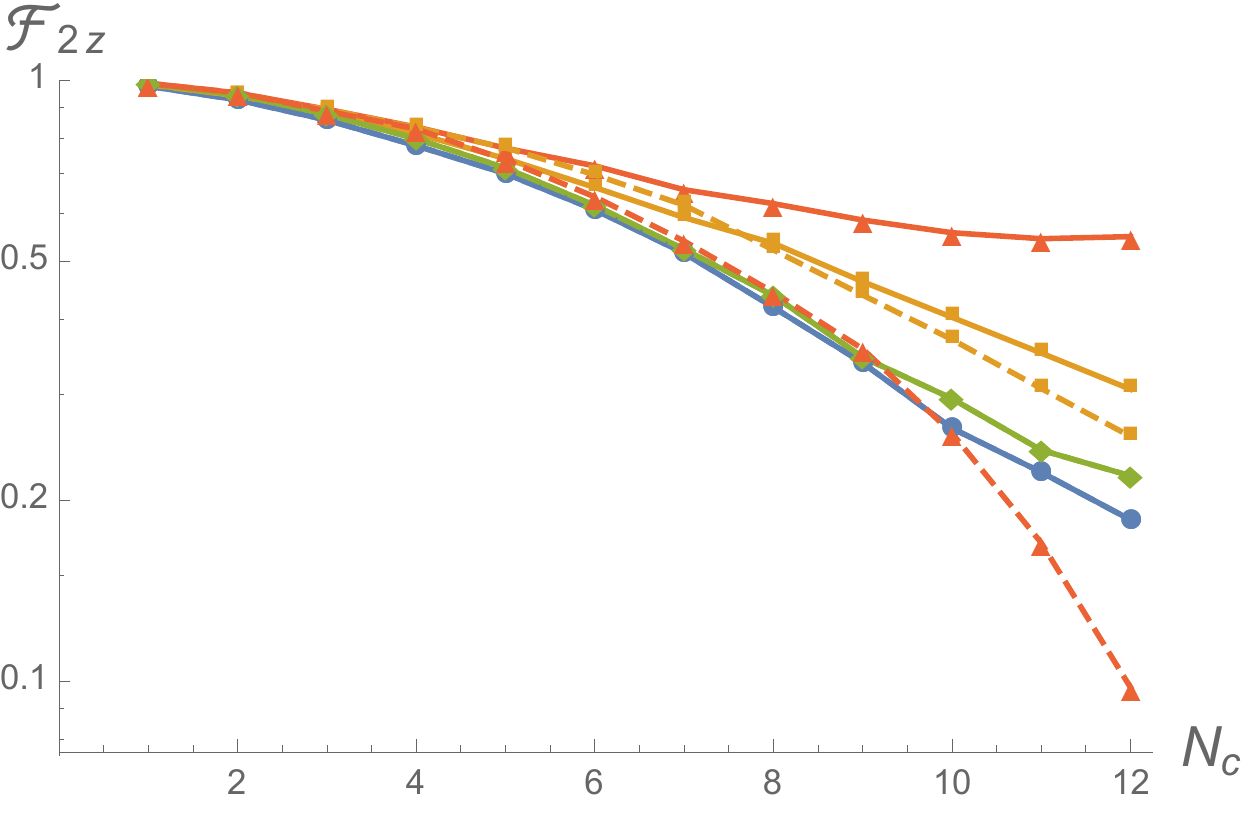}
\caption{Fidelity loss from one (top) or two (bottom) $z$ errors inserted at random spacetime points over 12 cycles in parametrization A, for $L=6,7,8,9$ (blue circles, gold squares, green diamonds, red triangles). At larger $L$, or more tellingly, higher numbers of photons added by the cavity interactions, the system's susceptibility to phase noise markedly decreases. The dashed lines for $L=7,9$ correspond to initializing the system with one additional photon in the qubits (3 total for $L=7$ and 4 total for $L=9$), which reduces the average number of photons added through interaction with the cavities; those instances' sensitivity to phase noise is substantially higher. A possible reason for this is discussed in the main text. As in the unitary protocol, single photon loss error sends $\mathcal{F}$ to zero.}\label{fidZfig}
\end{figure}

As in the unitary protocol, we find that a single photon addition or loss error leads to zero fidelity, though this was not guaranteed a priori in the noisy chain since photon number is not conserved. However, unlike in the unitary protocol, these events cannot be removed by post-selection, and so will directly reduce the observed fidelity in an experiment. We find that $z$ errors likewise reduce the fidelity, though as shown in FIG.~\ref{fidZfig} the extent to which one or two $z$ errors reduces $\mathcal{F}$ is much more variable, with the system displaying an apparent transition toward phase noise \textit{resilience} when the number of photons added by the cavities is more than $\sim1$. This is puzzling because, as seen earlier, other complexity-related observables such as entanglement, divergences from Porter-Thomas and IUR, and IPR, display similar behavior to the unitary case, and given this one would expect similar fragility to qubit phase noise in our noisily evolving chain.  

One possible reason for this could be a measurement effect from photon loss in the cavities-- as discussed earlier, a photon loss from a cavity projects the system's full wavefunction onto the subspace where a photon has been added or removed from the qubit chain via a very complex nonlocal operation, and that projection may decrease the resulting scrambling from a qubit $z$ error that occurred prior to it. If it is likely that a cavity-mediated photon addition occurs after the $z$ error has, then one would assume the fidelity loss from the $z$ error could be lower. As shown in FIG.~\ref{fidZfig}, simply initializing the system with one additional photon for $L=7$ and 9, which correspondingly reduces the average number of photons added by 20-30\%, is sufficient to eliminate the phase noise resilience of those instances, bolstering this interpretation. 

If this projection onto the action of nonlocal operators is indeed responsible for suppressing phase noise, one might naturally worry that it could lead to routes to efficient classical simulation, if the nonlocal operators themselves can be straightforwardly computed. We emphasize however that this should not be the case. As argued earlier, computing the appropriate matrix elements requires detailed knowledge of high-energy excited states (near the middle of the system's full spectrum) of an interacting system with disorder, and while we might be able to make predictions about instantaneous eigenstates near the ground state using perturbation theory or Arnoldi diagonalization, both methods break down once we go higher in the spectrum, necessitating the diagonalization of the full Hamiltonian. Even just focusing on the qubit subspace and ignoring the effect of double occupancies, the cost of doing so is $O \of{2^{2L}}$ space and $O \of{2^{3L}}$ time, making these operators impossible to compute in practice once the system gets reasonably large. Further, this argument likely applies to \textit{any} simulation method which attempts to eliminate the cavities by constructing new effective Lindblad operators for the qubits. For the parameters considered in this work, $\Omega_{QC} > \Gamma_C$, so the internal dynamics of the cavities cannot be ignored and they cannot be treated as purely Markovian noise sources. But even if this were not the case, when we include relatively sharp energy modulation of the matrix elements of a local creation or annihilation operator $a_i^\dagger$, the resulting transformed operator $\tilde{a}_i^\dagger$ is no longer sparse, requiring a cost $O \of{2^{2L}}$ to store it and $O \of{2^{3L}}$ operations to compute it. As we shall see below in the classical difficulty estimates section, this scaling is actually \textit{worse} than the time and memory costs of direct time evolution in a truncated basis, which does not involve any uncontrolled approximations.

Given realistic numbers, the fidelities achievable in this protocol are reasonably good. The previous unitary chain experiment reported fidelity reductions of approximately 5\% per qubit for state preparation and measurement (which was largely eliminated through post-selection), and 0.4\% per qubit per cycle for phase and control error accumulated during evolution. Assuming (a) no improvement in SPAM error and (b) no net increase in phase/control error due to the introduction of the sideband terms, 27 qubits evolved for 9 cycles would have an experimental fidelity of approximately 9.5\%, which is still good enough to clearly distinguish the contributions from quantum dynamics to the observed output, and an order of magnitude larger than typical fidelity targets for RQC. Since SPAM error well below 5\% has been realized in other experiments, it is reasonable to assume this could be brought down to 2\% with suitable hardware refinement, which would increase the fidelity to 22\%. Improvement in the per-cycle error is a trickier issue, as the protocol's apparent reduced sensitivity to phase noise would likely be balanced to some degree by the introduction of the sideband terms, which obviously bring with them additional error sources that would have to be carefully calibrated away.

\subsection{Fidelity versus number of cavity photons in simulation}

While the cavity photon populations are not measured in our protocol-- indeed, the cavities themselves are expected to be used to projectively measure the qubits, erasing any information about their own state-- they must be included in the system's Hilbert space for an accurate classical simulation. However, due to the fast loss rate and comparatively weak interaction between cavities and qubits, the actual photon populations in the cavities are expected to be low, and as a result substantial savings can be attained in classical simulation by truncating the maximum number of photons in the cavity Hilbert space. Doing so will reduce the fidelity relative to a full simulation including the entire cavity Hilbert space, but by precisely how much is a matter that must be estimated in numerical simulation. 

\begin{figure}
\includegraphics[width=3.25in]{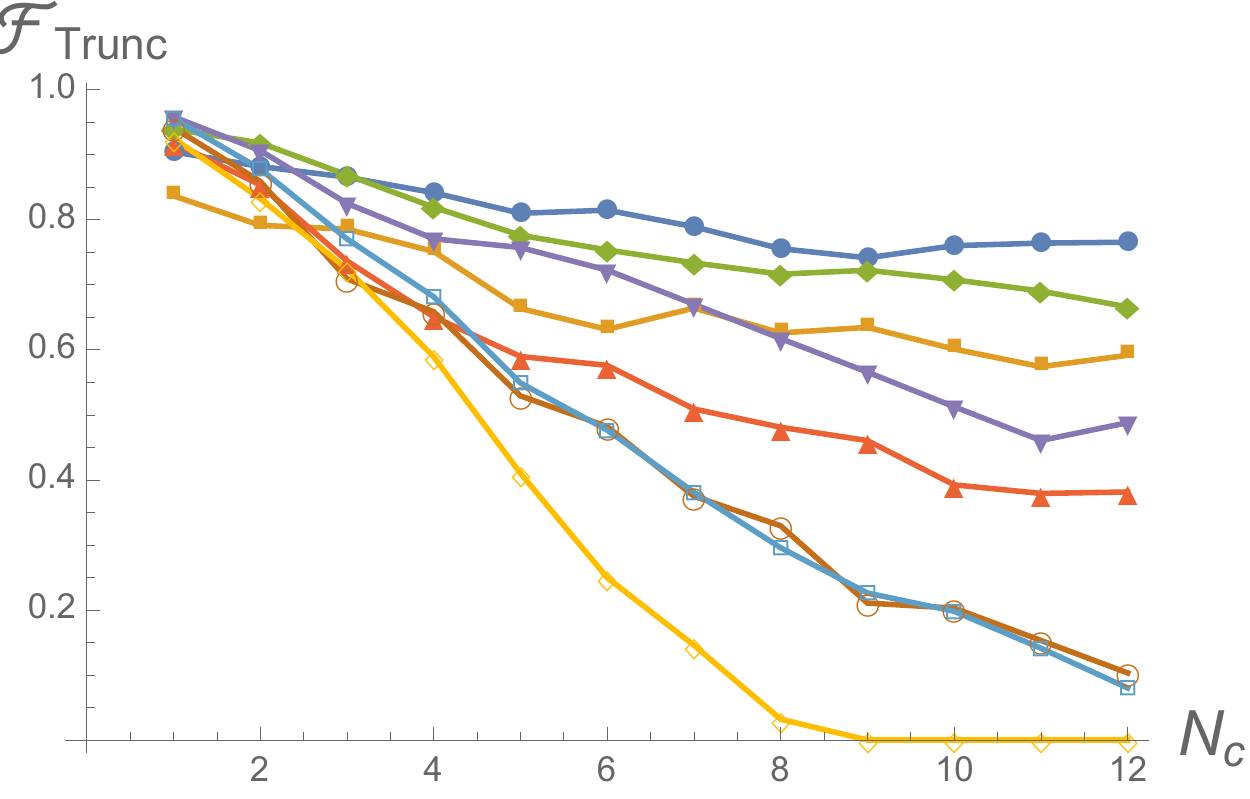}
\includegraphics[width=3.25in]{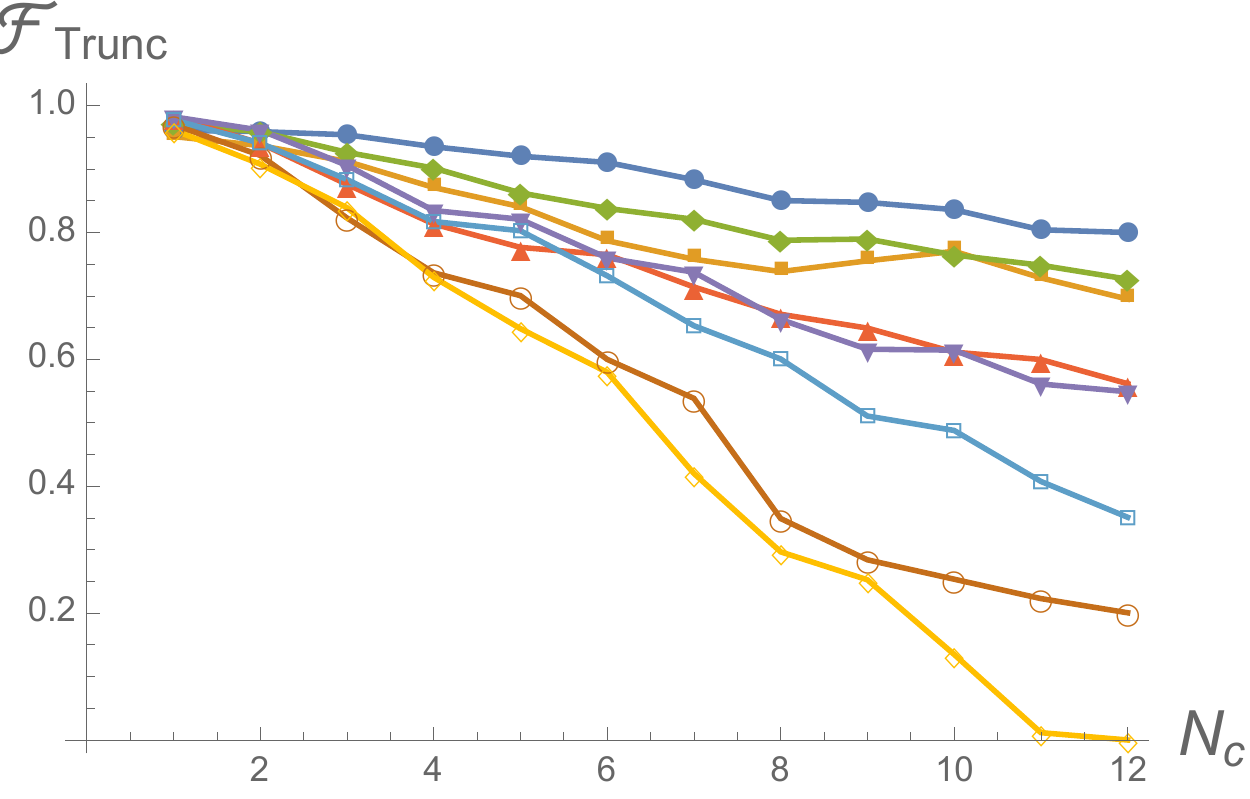}
\caption{Left to right: fidelity (defined in Eq.~\ref{defKLfid}) after restricting the system to have at most one photon in the cavity Hilbert space (in comparison to truncating it to at most two photons), for parametrizations A and B (top and bottom), for $L$ running from 4 to 11, using simulation parameters described in the text. These fidelities give a rough estimate of how many cavity photons need to be included in the cavity Hilbert space for an accurate simulation, and thus have important implications for the classical simulation difficulty of our protocol.}\label{fid12fig}
\end{figure}

In FIG.~\ref{fid12fig} we plot the fidelity loss from truncating from a maximum of two photons in the cavities (which we expect to be sufficient for the system sizes studied) to just one. These fidelity losses are important, since they can be used to estimate the classical simulation difficulty. As we will describe shortly, for methods which store the full evolving wavefunction, increasing the maximum number of cavity photons increases the size of the state and the time costs to evolve it. For methods which scale exponentially in entanglement, such as MPS or tensor network constructions, higher cavity photon populations increase the total explored Hilbert space and thus the maximum possible entanglement of the evolving state; in either case, higher cavity photon populations suggest a more complex classical simulation is necessary to accurately capture the system's evolution.

Having studied the output of our protocol in detail, we now turn to the question of the asymptotic classical difficulty to simulate it. We shall see that, due to the enlarged Hilbert space from including lossy cavities in the evolution, the threshold beyond which classical simulation is impossible should lie at substantially smaller system sizes than in the unitarily evolving chain upon which our protocol is based. We very roughly estimate that values of $L$ in the mid to high twenties are likely beyond the reach of near-term classical supercomputers, though we cannot rule out the possibility of more efficient simulation algorithms that would push this threshold higher.

\section{Classical difficulty estimates}

We now consider the projected difficulty of classically simulating the evolution in this circuit as $L$ becomes large. We assume throughout this section that the most efficient method is an average over quantum trajectories based on direct evolution of the system's full wavefunction (in an appropriately truncated basis). We offer no formal proof that a more efficient algorithm does not exist, but as we discuss below, exponentially growing entanglement means that matrix product methods are unlikely to provide a significant advantage over direct evolution, and the partitioning and decomposition methods used to simplify random quantum circuit simulations \cite{pednault2017breaking,boixo2017simulation,haner20170,chen201864,bouland2018quantum,chen2018classical,liwu2018,markov2018quantum} are likely not applicable to continuous time evolution under a varying $H \of{t}$, with or without noise. Further, the cost of those methods scales exponentially with gate depth, and given the large $g_{max}$, 6-8 cycles of evolution in our chain roughly corresponds to a depth of 42-56 in RQC (where each qubit experiences a CZ an average of once per two cycles). In other words, evolution in this system corresponds to a relatively deep quantum circuit, so any method which scales exponentially in gate depth will likely fail to accurately capture its evolution. We thus make the reasonable assumtion that direct wavefunction evolution will be the most efficient simulation method.

Proceeding from this assumption, we build on the estimates in \cite{neillroushan2017} through the following inclusions: a total transmon Hilbert space consisting of $O \of{C L}$ (for some small $C$) manifolds with a fixed number $M$ of photons in each, a resonator Hilbert space including up to $L/D$ photons across all the resonators (for some $D$ again depending on the details of the protocol) and a total of $N_t \of{L}$ trajectories that must be averaged over. We assume that attempting to precisely predict the probabilities $P_k$ for real quantum hardware would forbid us from employing the qubit subspace truncation used in this work; a more precise calculation would instead truncate the space of double and triple occupancies to a fixed number of bands.

\begin{figure}\includegraphics[width=3.25in]{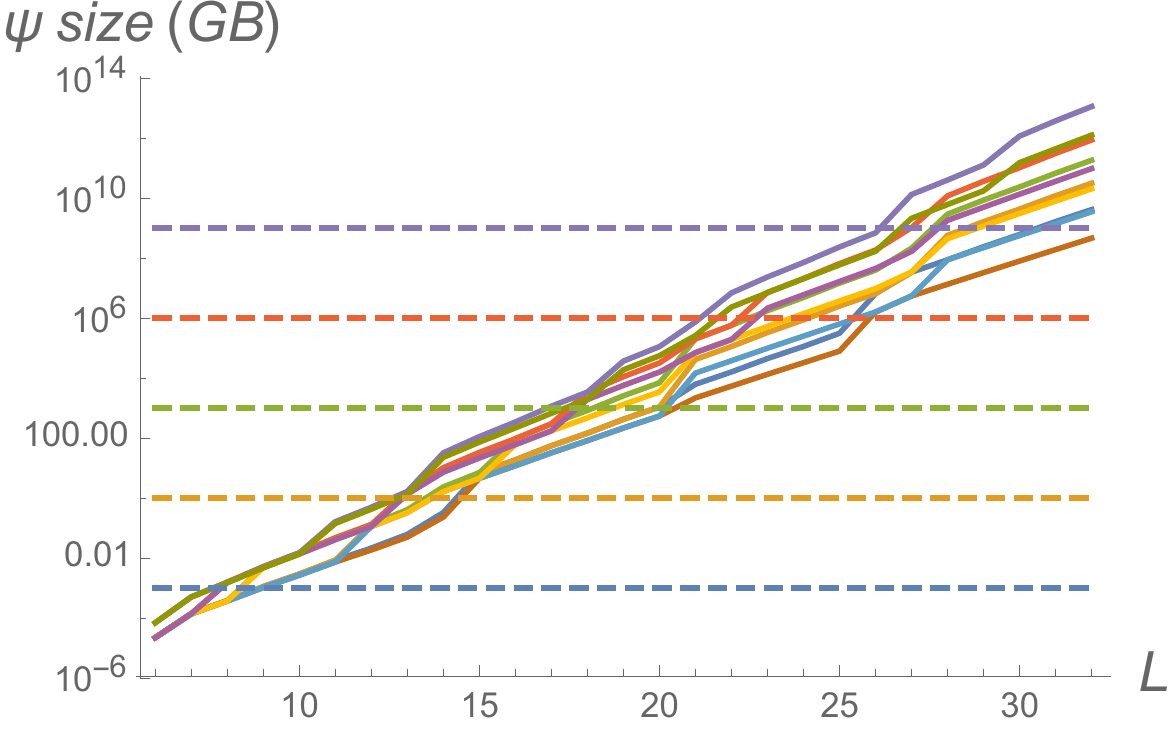}
\caption{Total wavefunction size vs. $L$, in gigabytes, for a range of possible Hilbert space truncations. As discussed in the text, we keep up to $\cuof{2,1}$ or $\cuof{3,2}$ $\{$doublons,triplons$\}$, a manifold of states with total qubit photon numbers in the range $\cuof{0.35L, 0.65L }$, and a maximum of $L/10$, $L/8$, $L/6$, $L/5$ or $L/4$ total photons in the cavities (all fractions are rounded to the nearest integer). Dashed lines correspond to a megabyte, gigabyte, terabyte, petabyte and exabyte, respectively.}\label{psisize}
\end{figure}

First, we estimate the memory requirements for estimating the $P_k$ based on direct wavefunction evolution. We plot a range of values, corresponding to simulations which keep up to $\cuof{2,1}$ or $\cuof{3,2}$ doublons/triplons in the qubit Hilbert space, and a manifold of states with total qubit photon numbers in the range $\cuof{0.35L, 0.65L }$ (fractions rounded to the nearest integer). We then tensor this with a cavity Hilbert space containing no double occupancies and a maximum of $L/10$, $L/8$, $L/6$, $L/5$ or $L/4$ total photons in the cavities, fractions again rounded to the nearest integer. The number of photons that need to be kept in the cavities depends on protocol details; as a very rough estimate, assuming that we need to include configurations with up to $L/D$ cavity photons (rounded to the nearest integer) to achieve reasonable fidelity, the results of FIG.~\ref{fid12fig} suggest that $D$ is in the range of 6 to 8. Assuming sixteen bytes per entry for double precision complex numbers, the total wavefunction storage sizes are plotted in FIG.~\ref{psisize}. The petabyte range is reached for $L$ between 21 and 26; the exabyte range between 27 and 34. 

Second, we estimate the time costs for this calculation. As argued in \cite{neillroushan2017}, the cost to unitarily evolve the full wavefunction for $L$ sites and a total time $T$ scales as $L^2 T N_H$, where $N_H$ is the Hilbert space size; this estimate comes from $O \of{L}$ terms in $H \of{t}$, a cost per sparse matrix-vector multiplication proportional to $N_H$, and a total number of matrix-vector multiplications proportional to $L T$, since the minimum timestep $dt$ scales as $1/L$. The cost to evaluate a single trajectory when noise is included scales similarly. Based on this and the empirical scaling of their simulations at smaller $L$, they provide a very rough estimate of 37 hours to fully evolve a 70 TB wavefunction over 1000 4th order Runge-Kutta steps on a 4096 node cluster with 1.2 GB/s per socket of node-to-node memory bandwidth. 

We expect that evolution with noise should take considerably longer. At the single trajectory level, the timestep $dt$ required for faithful simulation in a trajectory method is smaller than for unitary evolution, since in the unitary case errors in the wavefunction norm from each evolution step can be simply renormalized away, whereas in a noisy trajectory method decay of the wavefunction norm is tracked and used to determine when to randomly insert noise operations (see sec. III.D of \cite{daley2014}). More importantly, many trajectories must be evaluated for an accurate simulation. Comparing trajectory simulations with the full density matrix evolution for $L$ running from 4 to 8 led us to a very rough estimate that approximately $3 L \times N_c$ trajectories were needed to evolve an $L$-site system over $N_c$ cycles, with an output distribution that had an average K-L divergence from the exact result of 0.01 or less (we typically used $6 L^2$ trajectories in our simulations). Note that due to the nonlinearity intrinsic to how the K-L divergence is calculated, the K-L divergence from sampling a finite number of trajectories $N_t$ decreases as $1/N_t$, not $1/\sqrt{N_t}$ as in most other quantities. To see why this is, let $P_{k,ex}$ be the exact probability of obtaining bitstring $k$, and let $P_{k,N_{t}}$ be an approximation computed from $N_t$ trajectories. We can write $P_{k,N_{t}} = P_{k,ex} + \delta P_k$, where the individual $\delta P_k$ scale as $1/\sqrt{N_t}$ but due to normalization, $\sum_k \delta P_k = 0$ regardless of how few trajectories are sampled. If we then plug this into (\ref{defKL}) and expand the logarithm, the lowest order nonvanishing term is $\frac{1}{2} \sum_k \frac{ \of{ \delta P_k}^2}{P_{k,ex}}$, which is quadratic in the $\delta P_k$ and thus scales as $1/N_t$.

\begin{figure}
\includegraphics[width=3.25in]{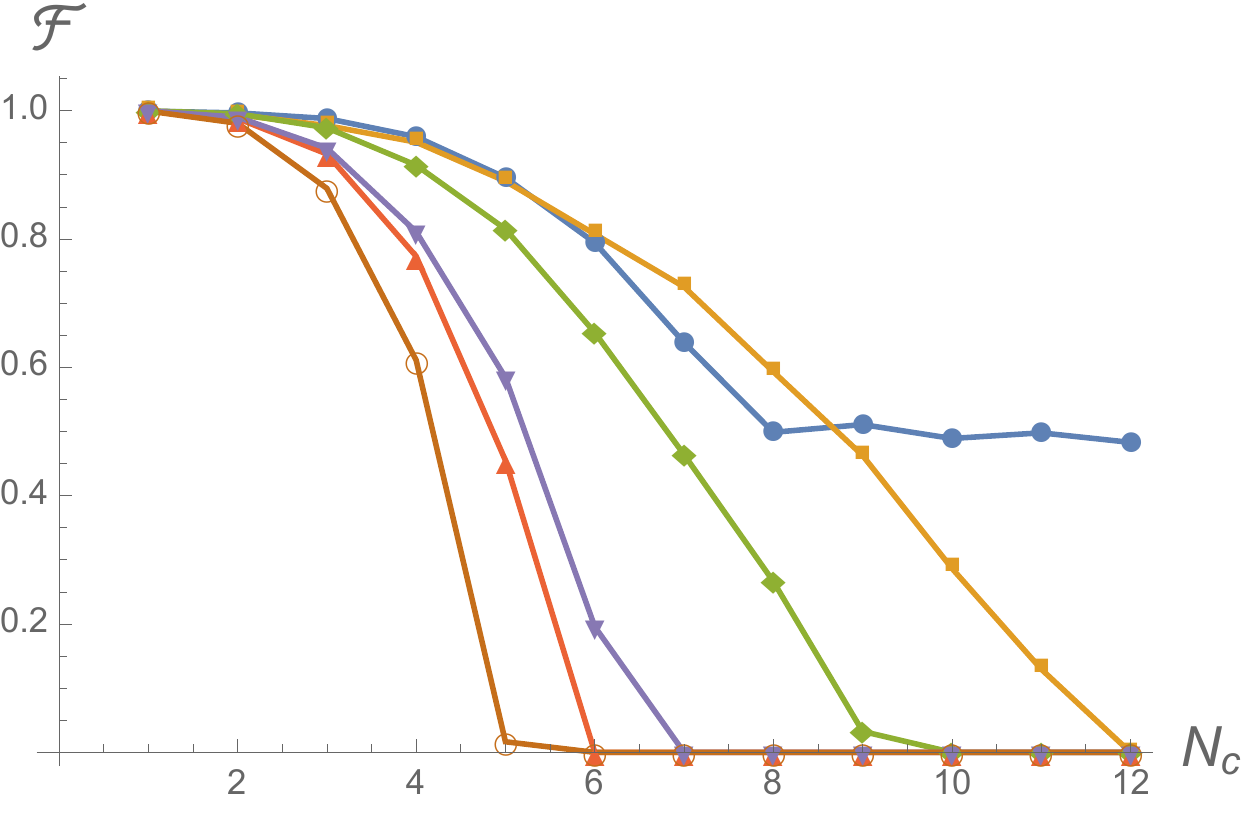}
\caption{Fidelity of a simulation with no cavity photon loss, compared to the full simulation, for parametrization A. For all $L > 4$ the results of the two methods rapidly diverge, indicating that our noisy protocol cannot be simulated with noise-free methods. This implies that many trajectories will have to be sampled to obtain an accurate result, significantly increasing the runtime of a classical simulation.}\label{Uvsfullfig}
\end{figure}

This consideration aside, the estimate $3 L \times N_c$ stretches into the hundreds when the wavefunction approaches the PB scale, and suggests that runtime may ultimately prove to be the limiting factor in an accurate simulation, given that parallelization of the trajectories would quickly become memory-limited even on the largest current supercomputers. Note that, given experimental error the infidelity of the real experiment would likely be much worse than this so one could get away with sampling fewer trajectories, though given the need for a smaller $dt$ and other complications we still expect that runtime should be a significant bottleneck. For further evidence that the need to average over trajectories is unavoidable, we also simulated evolution with the cavity loss rate $\Gamma_C$ set to zero (but still including the truncated cavity Hilbert space and the qubit-cavity interaction terms), and compared the result of that perfectly unitary evolution to the full simulation. As shown in FIG.~\ref{Uvsfullfig}, the fidelity drops to zero within just a few cycles, confirming that the noise processes cannot be ignored in classical simulation.

\subsection{Matrix product state methods}
An obvious possible objection to the above estimates is that matrix product state (MPS) methods, which have time and memory costs that scale exponentially in the system's total entanglement and not Hilbert space size, may prove more efficient. Given the many successes of MPS methods in other contexts \cite{orus2014practical}, it is natural to ask whether or not they could simplify the simulation of our noisy chain. Note that these questions likely do not apply in a 2d implementation, where we expect most of our claims about entanglement and complexity to still hold, but MPS or tensor network methods are significantly less effective.

\begin{figure}
\includegraphics[width=3.25in]{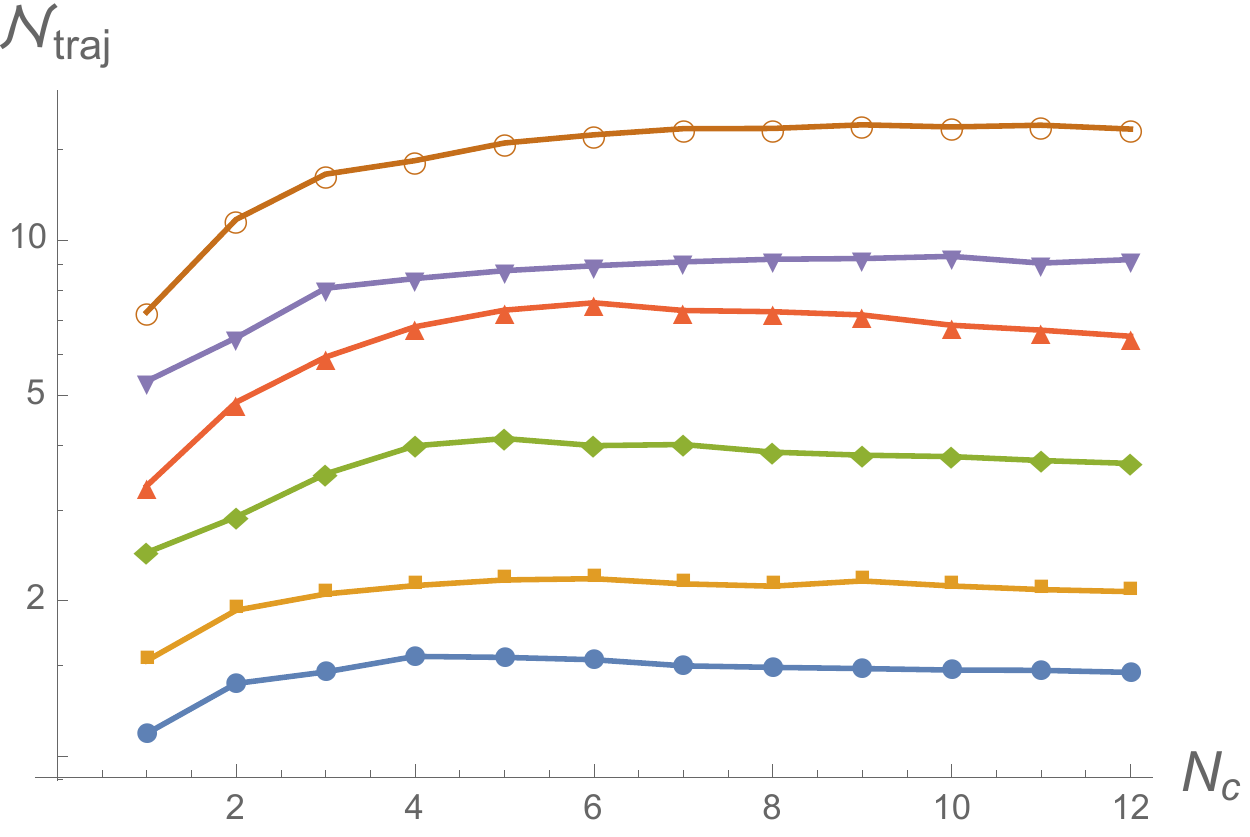}
\includegraphics[width=3.25in]{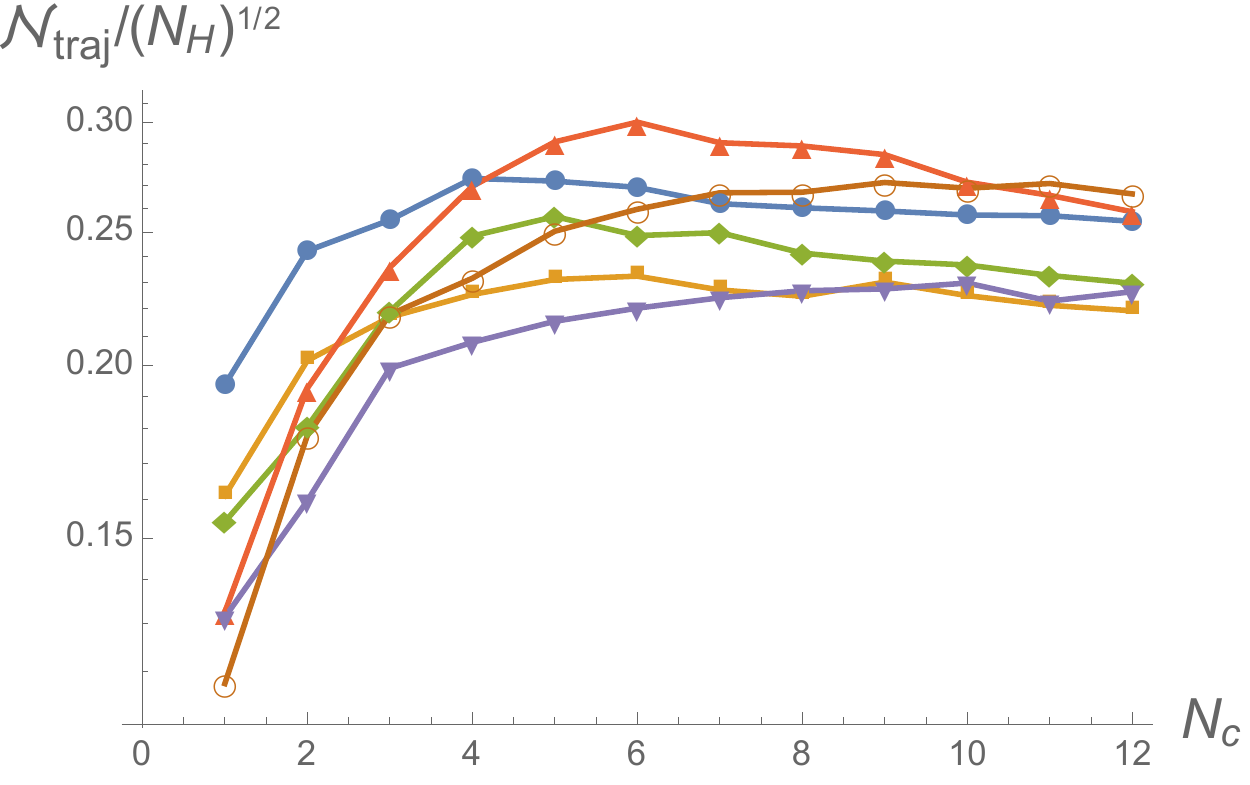}
\caption{Average \textit{per-trajectory} entanglement negativity (top), and the same quantities plotted as a ratio of $\mathcal{N}$ to the square root of the total Hilbert space size, $\sqrt{N_H}$. Unlike the negativity of the full simulation shown in FIG.~\ref{negfig} (which is averaged over many random trajectories), the per-trajectory negativity does not appreciably decay at long times, since the system can rapidly re-entangle after a cavity photon loss. As discussed in the text, this result suggests that matrix product state based methods for simulation evolution in our protocol will not be efficient and likely will exhibit worse scaling than direct wavefunction evolution.}\label{negpertrajfig}
\end{figure}

Assuming that the volume entanglement scaling we observed in our simulations persists to larger $L$, we expect that this should not be the case. The memory cost to store an MPS wavefunction over $L$ sites with negativity $\mathcal{N}$ is approximately $4 L d \mathcal{N}^2$ complex numbers, where $\mathcal{N}_{max} \simeq \sqrt{N_H}/2$, $d$ is the local dimension of each site and $N_H$ is the size of the full Hilbert space. Treating each qubit-cavity pair as a composite object and including states up through $\ket{3}$ gives $d=8$ for our chain. The cost to unitarily time evolve such a state is higher by a factor proportional to $d^2 \mathcal{N}$. MPS methods can be used in noisy systems, through for example the quantum trajectory methods in \cite{daleytaylor2009,bonnes2014superoperators}. However, the memory cost of a quantum trajectory simulation using MPS states is based on the negativity of a typical trajectory, and not the averaged negativity, and this can be substantially higher since the system rapidly re-entangles after a photon loss. In FIG.~\ref{negpertrajfig} we calculate the average per-trajectory negativity in our chain, and show that, unlike the full dynamics averaged over random quantum jumps, it does not decay with time and remains an $O \of{1}$ fraction of $\mathcal{N}_{max}$. We attribute this difference to the system rapidly re-entangling after a photon loss; unlike the full system the entanglement of the evolving state in a single trajectory is not suppressed since we are not averaging over different locations (in time and space) for the photon loss operator insertions. 

Consider also the decay of entanglement vs. average number of photon losses, discussed earlier in this work, which would be relevant to methods which scale with the average negativity and not the per-trajectory negativity. We found that after a sufficiently long time, for an average of $p$ cavity photon losses the bipartite negativity scales as $\mathcal{N} \of{p} \simeq \mathcal{N}_0 e^{- c_{L} p}$, where $\mathcal{N}_0 \propto \mathcal{N}_{max}$ and $c_L$ depends on $L$ and protocol details, and is generally close to but slightly less than 1. Since $p$ scales linearly with $L$ and the number of cycles (see FIG.~\ref{nqnlfig}), extrapolating to $L=27$ at 8 cycles gives $p \simeq 2.7$ for parametrization A and $p \simeq 2.25$ for parametrization B, the resulting entanglement loss assuming $c_L = 1$ should give a negativity equivalent to that of a volume-entangled system with between six and eight fewer total qubit degrees of freedom (e.g. a total system Hilbert space smaller by a factor between 64 and 256). As an alternative estimate, we took the negativity measured at 7, 8 or 9 cycles for each protocol as a function of $L$, and fit that to $A \mathcal{N}_{max} \of{L} 2^{-d L}$, where $\mathcal{N}_{max}$ is the maximum possible negativity assuming at most two photons in the cavities and $A$ and $d$ are fitting parameters; those fits returned values of $d$ ranging from 0.08 to 0.13, and thus predict a negativity equivalent to true volume entanglement with 4-7 fewer qubits for $L=27$ at 8 cycles, a nearly identical range. This is significant, but when the additional time evolution cost of $d^2 \mathcal{N}$ is taken into account we expect that MPS methods should still be substantially less efficient than direct Schrodinger evolution averaged over trajectories. These results suggest that MPS simulation methods will be more expensive than the full wavefunction evolution, particularly given that runtime could prove to be a bottleneck before memory does, due to the large number of trajectories involved in a faithful simulation.

All that said, one could attempt a matrix product simulation where the total entanglement is bounded to reduce computational difficulty. This would reduce the simulation fidelity relative to a full wavefunction evolution, perhaps to an acceptable degree (e.g. below the expected fidelity loss from various error sources in the real experiment). The details and scaling of such calculations are beyond the scope of this paper, though the possibility deserves further exploration.

\section{Conclusions and Outlook}

In this work, we presented a deceptively simple modification to a leading quantum sampling problem-- weak but resonant coupling to lossy cavities-- and showed that it leads to dramatic changes in the quantum dynamics. By considering a wide range of metrics in direct numerical simulation, we showed that features suggesting classical intractability, including volume-law entanglement and an output distribution consistent with quantum chaotic evolution at intermediate times, persist despite the presence of strong noise in the system. These results suggest that quantum sampling problems including noise in their definition can still be extremely difficult to solve with classical machines, and are thus potentially good candidates for demonstrating a quantum advantage in near-term hardware. This is doubly true for superconducting platforms, where lossy elements in the form of readout cavities are already present for qubit measurement, and involving them in the state's evolution can greatly increase the \textit{quantum simulation} complexity without increasing the \textit{hardware} complexity of the implementation. These methods, or variations of them, likely represent the most difficult simulation problem that can be practically engineered with a given number of transmon qubits (and associated measurement cavities).

For a variety of reasons, the basic protocol in this work, and the parameters used in its numerical simulation, were closely tied to the previously reported gmon chain experiment. However, the fundamental mechanism-- pulses of delocalized evolution through tunneling terms combined with much weaker, resonant, driven interactions coupling the primary system to a lossy auxiliary one-- is fairly generic, and we have no doubts that variations of it would produce similar results. That said, when compared to unitary sampling problems such as the isolated Bose-Hubbard chain or RQC, families of dissipative protocols can be qualitatively more sensitive to changes in protocol details (e.g. what classes of operator to use in $H_P$, the choice of which sideband terms to employ, the choice of resonance energies for the lossy objects, etc), and some choices may lead to results which can be efficiently reproduced classically. For example, simply alternating the qubit-cavity and qubit-coupler pulses, rather than operating them simultaneously as done in this work, can lead to a situation where the qubits are repeatedly subjected to effective local measurements, which disentangle the state and open the door to efficient classical simulations. Further, it strikes us as unlikely on general grounds for experimentally realistic protocols with substantial dissipative elements to exhibit chaotic behavior at arbitrarily long times, though the intermediate-time behavior of the protocols considered in this work certainly appears to be, and the time scale of quantum chaos can be increased by reducing the loss rate of the dissipative elements.

Finally, the techniques described in this work allow for an intriguing future application: the simulation of \textit{thermal} many-body states using superconducting circuits. Multiple previous proposals \cite{hafeziadhikari2015,shabanineven2015} have argued that a thermal bath can be simulated in interacting photon systems using suitably complex bath structures, though when these constructions are combined with intrinsic qubit noise the character of the resulting steady state, and its effective temperature, remain an open question. However, methods developed in studying cold atoms \cite{zhou2011universal} allow the system's temperature to be extracted from local density fluctuations in the presence of a slowly varying potential (even if the underlying microscopic Hamiltonian is not known), so sufficiently large circuits could be used to probe the thermodynamics of novel interacting boson systems. In cases where the system is small or analytically simple enough to permit a classical solution, this measure could be further bolstered by directly comparing the observed output distribution to a theoretical model using the K-L divergence or a similar sampling metric. These approaches could greatly expand the space of models that can be probed in analog quantum simulation.

\section{Acknowledgements}

Eliot Kapit's research is supported by the National Science Foundation via grant PHY-1653820, and by Google, Inc. He would like to thank L. Carr, D. Jaschke, and V. Oganesyan for useful discussions related to this work.

\bibliography{fullbib}

\end{document}